\documentclass[a4paper,twocolumn,showpacs,floatfix,prd,nofootinbib]{revtex4-1}
\usepackage{graphicx}
\usepackage{epsf}
\usepackage{epsfig}
\usepackage{amssymb,amsmath}
\usepackage[usenames]{color}
\usepackage{times}
\usepackage[colorlinks=true,linktocpage=false,citecolor=blue,linkcolor=blue,urlcolor=blue]{hyperref}
\usepackage{enumerate}
\usepackage{booktabs,dcolumn}
\usepackage{pstricks}
\usepackage{rotating}
\newcolumntype{d}[1]{D{.}{.}{#1}}

\def\nn{\nonumber}
\def\be{\begin{equation}}
\def\ee{\end{equation}}
\def\beq{\begin{eqnarray}}
\def\eeq{\end{eqnarray}}
\def\bi{\begin{itemize}}
\def\ei{\end{itemize}}
\def\ben{\begin{enumerate}}
\def\een{\end{enumerate}}

\begin{document}

\author{Francesco Pannarale}%
\email{francesco.pannarale@aei.mpg.de}%
\affiliation{%
  Max-Planck-Institut f{\"u}r Gravitationsphysik, Albert Einstein
  Institut, Potsdam 14476, Germany\\
  and School of Physics and Astronomy, Cardiff University, The Parade,
  Cardiff CF24 3AA, United Kingdom%
}

\title{\bf Black hole remnant of black hole-neutron star coalescing
  binaries}

\begin{abstract}
  We present a model for determining the dimensionless spin parameter
  and mass of the black hole remnant of black hole-neutron star
  mergers with parallel orbital angular momentum and initial black
  hole spin. This approach is based on the Buonanno, Kidder, and
  Lehner method for binary black holes, and it is successfully tested
  against the results of numerical-relativity simulations: the
  dimensionless spin parameter is predicted with absolute error
  $\lesssim 0.02$, whereas the relative error on the final mass is
  $\lesssim 2$\%, its distribution in the tests being pronouncedly
  peaked at $1$\%. Our approach and the fit to the torus remnant mass
  reported in~\cite{Foucart2012} thus constitute an easy-to-use
  analytical model that accurately describes the remnant of black
  hole-neutron star mergers. The space of parameters consisting of the
  binary mass ratio, the initial black hole spin, and the neutron star
  mass and equation of state is investigated. We provide indirect
  support to the cosmic censorship conjecture for black hole remnants
  of black hole-neutron star mergers. We show that the presence of a
  neutron star affects the quasinormal mode frequency of the black
  hole remnant, thus suggesting that the ringdown epoch of the
  gravitational wave signal may virtually be used to (1) distinguish
  black hole-black hole from black hole-neutron star mergers and to
  (2) constrain the neutron star equation of state.
\end{abstract}

\maketitle

%%%%%%%%%%%%%%%%%%%%%%%%%%%%%%%%%%%%%%%%%%%%%%%%%%%%%%%%%%%%%%%%%%%%%%%%%%%%%%%
\section{Introduction}
%%%%%%%%%%%%%%%%%%%%%%%%%%%%%%%%%%%%%%%%%%%%%%%%%%%%%%%%%%%%%%%%%%%%%%%%%%%%%%%
Once a black hole-neutron star (BH-NS) binary is formed, gravitational
radiation reaction gradually reduces its orbital separation until the
two companions merge and leave behind a remnant consisting of a black
hole and, possibly, a hot, massive accretion torus surrounding
it~\cite{ShibataTaniguchilrr-2011-6}. BH-NS binaries have not been
observed yet; population synthesis studies, however, suggest that the
coalescence of BH-NS systems is likely to occur frequently in the
Hubble volume, thus making theoretical studies on the evolution and
final state of BH-NS mergers relevant~\cite{Kalogera2007,
  Belczynski07, Belczynski08, Oshaughnessy2008, Abadie:2010}. Interest
in these systems arises from the fact that they are among the most
promising sources for gravitational wave (GW) detectors --- such as
LIGO~\cite{ligowebpage}, Virgo~\cite{virgowebpage},
KAGRA~\cite{kagrawebpage}, and the Einstein
Telescope~\cite{Punturo:2010} --- and that they are promising
candidates as progenitors of (a fraction of) short-hard gamma-ray
bursts~\cite{Nakar:2007yr, Berger2011}. Further, as NSs in these
systems undergo strong tidal deformations, observing GW and/or
electromagnetic signals emitted by BH-NS binaries could help shed
light on the equation of state (EOS) of matter at supranuclear
densities, which is currently unknown~\cite{Vallisneri00,
  Ferrari:2009bw, Pannarale2011, Lackey2012}. Finally, comprehending
the fate of the material possibly ejected by BH-NS binaries after the
NS tidal disruption is relevant in interpreting the observed
abundances of the heavy elements that are formed by rapid neutron
capture in $r$-processes~\cite{Lattimer74}. These outflows may
additionally be observable due to the radioactive decays triggered by
the formation of heavy isotopes, i.e.~``kilonovas,'' or due to the
shock they would generate when hitting interstellar medium of
sufficiently high density~\cite{Metzger2012}.

To achieve a full understanding of BH-NS merger events and their
physics, numerical-relativity simulations are required. These will
ultimately have to include adequate and accurate treatments of general
relativity, relativistic (magneto)hydrodynamics, the microphysical
EOS, NS crust physics, thermal effects, and nuclear physics
reactions. Numerical quasiequilibrium studies~\cite{Taniguchi05,
  Grandclement06, Faber06a, Faber05, Tsokaros2007, Taniguchi07,
  Taniguchi:2008a} and dynamical
simulations~\cite{Janka99a,Ruffert99b, Kluzniak99c, Rosswog05,
  Loeffler06a, Sopuerta:2006bw, Shibata06d, ShibataUryu:2007,
  Etienne2007b, ShibataTaniguchi2008, Rantsiou2008, Duez:2008rb,
  Etienne:2008re, Shibata:2009cn, Duez09, Kyutoku2010, Chawla:2010sw,
  Foucart2010, Kyutoku2011, Kyutoku2011err, Foucart2011, Lackey2012,
  Etienne2012, Shibata2012err, Etienne2012b, Deaton2013, Foucart2013a,
  Kyutoku2013, Foucart2013b, Lovelace2013, Paschalidis2013b} of mixed
binary mergers made considerable progress in the last few
years. Despite the fact that simulating BH-NS mergers is now possible,
these simulations remain nevertheless both challenging and
computationally intensive. This has motivated the parallel development
of pseudo-Newtonian BH-NS calculations, e.g.~\cite{Ruffert2010}, and
analytical approaches focusing on specific physical aspects of the
problem, e.g.~\cite{Shibata96, Wiggins00, Vallisneri00, Berti08,
  Hanna2008, Ferrari09, Ferrari:2009bw, Damour:2009wj, Pannarale2010,
  Pannarale2011, Foucart2012}. Studies of these kinds benefit from
their low computational costs which allow them to shed light on
questions that cannot be currently addressed with numerical
simulations and to provide insight on what happens when the large
space of parameters of BH-NS binaries is spanned. They may, in turn,
aid numerical relativity by suggesting cases that are particularly
interesting to simulate, and by providing information to exploit
within the simulations themselves.

In this paper we focus on predicting the spin parameter and mass of
the BH remnant of BH-NS coalescing binaries by using a semianalytical
approach. While this problem has a fairly long history in the case of
coalescing binary black holes~\cite{Buonanno00a, Damour:2001tu,
  Hughes:2002ei, Buonanno:06cd, Campanelli:2006gf, Campanelli:2006uy,
  Campanelli:2006vp, DamourNagar:07a, Tichy:2007gso, Buonanno:07b,
  Rezzolla-etal-2007, BoyleKesdenNissanke:07, Kesden:2008,
  Marronetti:2007wz, Rezzolla-etal-2007b, Rezzolla-etal-2007c,
  Gergely:08, Rezzolla:2008sd, Barausse:2009uz, Kesden2010,
  Barausse2012b}, no attempt beyond numerical-relativity simulations
has yet been made to tackle it in the case of BH-NS mergers. The
approach we present and discuss is based on the work of Buonanno,
Kidder, and Lehner (BKL) on estimating the final BH spin of a
coalescing binary BH with arbitrary initial masses and
spins~\cite{Buonanno:07b}. We choose this simple, phenomenological
model as a starting point because it provides good physical insight,
and because it is straightforward to modify and extend. Our method may
indeed be seen as a generalization of the BKL model to the case in
which the lower mass BH is replaced with a NS. For the time being,
however, it is restricted to systems in which the BH spin direction is
parallel to the orbital angular momentum direction. The closed
expression we determine for the final spin parameter automatically
yields an estimate of the mass of the BH remnant by means of a method
similar to the starting point of Barausse, Morozova, and Rezzolla's
calculations on the mass radiated by binary BHs~\cite{Barausse2012b},
but with modifications inspired, once again,
by~\cite{Buonanno:07b}. The key equations of our approach are
Eqs.\,(\ref{eq:model-Mf}), (\ref{eq:model-imp})-(\ref{eq:fbridge})
and, despite the mathematical complexity of the mixed binary
coalescence problem, our method enables us to reproduce the results of
numerical-relativity simulations with reasonable accuracy.

The paper is organized as follows. In Sec.\,\ref{sec:BKL} we review
the BKL approach for binary BHs. In Sec.\,\ref{sec:model} we propose
an extension of this method in order to predict the spin parameter and
mass of BH remnants of BH-NS mergers --- Eqs.\,(\ref{eq:model-Mf}),
(\ref{eq:model-imp})-(\ref{eq:fbridge}) --- and successfully test it
against available numerical-relativity data. In
Sec.\,\ref{sec:results} we gather the results obtained by
systematically varying the binary mass ratio, the initial BH spin
parameter, and the NS mass and EOS. First, we provide indirect support
to the cosmic censorship conjecture and suggest particularly
interesting cases to explore with numerical simulations in this
context (Sec.\,\ref{sec:maxaf}). Then, we show that the NS EOS may
leave an imprint on the BH remnant in terms of its final spin and mass
(Sec.\,\ref{sec:QNMs}). This suggests the idea of inferring the
presence of the NS and of constraining its EOS from the ringdown of
the BH remnant. Finally, in Sec.\,\ref{sec:conclusions}, we draw our
conclusions and collect our remarks.

%%%%%%%%%%%%%%%%%%%%%%%%%%%%%%%%%%%%%%%%%%%%%%%%%%%%%%%%%%%%%%%%%%%%%%%%%%%%%%%
\section{The BKL Formula}\label{sec:BKL}
%%%%%%%%%%%%%%%%%%%%%%%%%%%%%%%%%%%%%%%%%%%%%%%%%%%%%%%%%%%%%%%%%%%%%%%%%%%%%%%
The BKL approach to estimate the final spin of BH-BH
mergers~\cite{Buonanno:07b} starts by considering an initial reference
state with two widely separated black holes approximated as two Kerr
black holes having masses $\{M_1,M_2\}$ and dimensionless spin
parameters\footnote{In~\cite{Buonanno:07b} $a$'s have the dimensions
  of a mass, while they are dimensionless throughout this paper.}
$\{a_1,a_2\}$. The case of the BKL approach that we will extend in
order to describe BH-NS binaries is that of BH binary systems, the
orbits of which stay within a unique plane, referred to as the
equatorial plane; in such cases, the orbital angular momentum and the
individual spins of the BHs are orthogonal to the equatorial
plane. The spin parameter of the BH remnant $a_\text{f}$ is obtained
in terms of the initial configuration of the system by a
phenomenological approach that relies on the following two
observations based on intuitive arguments, on post-Newtonian and
perturbative calculations for the inspiral and ringdown, and on
numerical simulations of the merger:
\ben[~~~~(1)]
\item The system evolves quasiadiabatically during the inspiral phase.
\item The total mass and angular momentum of the system change only by
  a small amount during the merger and ringdown phases.
\een
Further, the BKL expression for $a_\text{f}$ is derived from first
principles once the following assumptions are made:
\ben[~~~~(1)]
\item The mass of the system is conserved to first order, so that the
  final BH has a total mass $M = M_1 + M_2$.
\item The magnitude of the individual BH spins remains constant, and
  their contribution to the final total angular momentum is determined
  by the their initial values.
\item The system radiates much of its angular momentum in the long
  inspiral stage until it reaches the innermost stable circular orbit
  (ISCO), when the dynamics quickly leads to the merger of the two
  BHs. Given that the radiation of energy and angular momentum during
  the merger is small with respect to the mass and angular momentum of
  the system, the contribution of the orbital angular momentum to the
  angular momentum of the BH remnant is estimated by considering the
  orbital angular momentum of a test particle orbiting a Kerr BH, with
  spin parameter equal to that of the \emph{final} BH, at the ISCO.
\een
All these assumptions are combined in the following formula expressing
the dimensionless spin parameter of the final BH:
\begin{align}
\label{eq:BKL}
a_\text{f} = \frac{a_1M_1^2 + a_2M_2^2 +
  l_z(\bar{r}_\text{ISCO,f},a_\text{f})M_1M_2}{M^2}\,,
\end{align}
where $l_z(\bar{r}_\text{ISCO,f},a_\text{f})$ is the orbital angular
momentum per unit mass of a test particle orbiting the BH remnant at
the ISCO, and where we introduced the notation $\bar{r}=r/M$ for the
(dimensionless) Boyer-Lindquist radial coordinate.

We recall that for equatorial orbits around a Kerr BH of spin
parameter $a$,
\begin{align}
\label{eq:lz}
l_z(\bar{r},a) &= \pm \frac{\bar{r}^2\mp
  2a\sqrt{\bar{r}}+a^2}{\sqrt{\bar{r}}(\bar{r}^2-3\bar{r}\pm
  2a\sqrt{\bar{r}})^{1/2}},
\end{align}
and that the orbital separation at the ISCO is given by
\begin{align}
  \bar{r}_\text{ISCO} &= [3+Z_2\mp\sqrt{(3-Z_1)(3+Z_1+2Z_2)}]\nn\\
  Z_1 &= 1 +
  (1-a^2)^{1/3}\left[(1+a)^{1/3}+(1-a)^{1/3}\right]\nn\\
  Z_2 &= \sqrt{3a^2+Z_1^2}\,,
\end{align}
where upper/lower signs hold for co/counter-rotating
orbits. Throughout the paper we will use the symbols
$\bar{r}_\text{ISCO,i}$ and $\bar{r}_\text{ISCO,f}$ to denote
$\bar{r}_\text{ISCO}$ calculated for the initial and final BH spin
parameter, respectively. In the following, we will also be using the
energy per unit mass $e$ of a test particle orbiting a BH. It may be
expressed as
\begin{align}
\label{eq:e}
e(\bar{r},a) &= \frac{\bar{r}^2-2\bar{r}\pm
  a\sqrt{\bar{r}}}{\bar{r}(\bar{r}^2-3\bar{r}\pm
    2a\sqrt{\bar{r}})^{1/2}}
\end{align}
for Kerr equatorial orbits.

%%%%%%%%%%%%%%%%%%%%%%%%%%%%%%%%%%%%%%%%%%%%%%%%%%%%%%%%%%%%%%%%%%%%%%%%%%%%%%%
\section{A Model for BH-NS Mergers}\label{sec:model}
%%%%%%%%%%%%%%%%%%%%%%%%%%%%%%%%%%%%%%%%%%%%%%%%%%%%%%%%%%%%%%%%%%%%%%%%%%%%%%%
When modifying Eq.\,(\ref{eq:BKL}) in order to describe BH-NS systems,
the first step is to set the initial spin angular momentum of the NS
to zero since (1) this is believed to be a reliable approximation of
astrophysically realistic systems~\cite{Bildsten92,Kochanek92} and (2)
this was done in all BH-NS merger numerical simulations so far and we
use these as test cases to assess the validity of our model. Adapting
the notation in Eq.\,(\ref{eq:BKL}) to BH-NS binaries, we now have
\begin{align}
\label{eq:step1}
a_\text{f} = \frac{a_\text{i}M_\text{BH}^2 +
  l_z(\bar{r}_\text{ISCO,f},a_\text{f})M_\text{BH}M_\text{NS}}{M^2}\,.
\end{align}

In the case of disruptive BH-NS mergers, an accretion torus
surrounding the BH remnant may be formed and one must thus drop
assumption $1$ of the BKL approach and adequately modify
Eq.\,(\ref{eq:step1}) to take this possibility into account. This is
done by: (1) replacing the term
$l_z(\bar{r}_\text{ISCO,f},a_\text{f})M_\text{BH}M_\text{NS}$ in the
numerator with
$l_z(\bar{r}_\text{ISCO,f},a_\text{f})M_\text{BH}(M_\text{NS}-M_\text{b,torus})$,
and by (2) replacing $M$ with
$M-e(\bar{r}_\text{ISCO,f},a_\text{f})M_\text{b,torus}$ in the
denominator, where $M_\text{b,torus}$ is the baryonic mass of the
torus remnant. The former/latter replacement expresses the lack of
angular momentum/mass accretion onto the BH due to the formation of
the torus.\footnote{We are introducing the approximation
  $L_{z,\text{torus}}\equiv
  M_\text{BH}\int_\text{torus}d(l_zm_\text{b})\approx
  M_\text{BH}l_z(\bar{r}_\text{ISCO,f},a_\text{f})\int_\text{torus}d(m_\text{b})=M_\text{BH}l_z(\bar{r}_\text{ISCO,f},a_\text{f})M_\text{b,torus}$,
  and the approximation $E_\text{torus}\equiv
  \int_\text{torus}d(em_\text{b})\approx
  e(\bar{r}_\text{ISCO,f},a_\text{f})\int_\text{torus}d(m_\text{b})=e(\bar{r}_\text{ISCO,f},a_\text{f})M_\text{b,torus}$.}
In the case of no torus formation, $M_\text{b,torus}=0$ and full
accretion of both mass and angular momentum onto the BH is
achieved. Our formula now reads
\begin{align}
\label{eq:step2}
a_\text{f} = \frac{a_\text{i}M_\text{BH}^2 +
  l_z(\bar{r}_\text{ISCO,f},a_\text{f})M_\text{BH}(M_\text{NS}-M_\text{b,torus})}{[M-e(\bar{r}_\text{ISCO,f},a_\text{f})M_\text{b,torus}]^2}\,,
\end{align}
where we once more emphasize that $e$ and $l_z$ are calculated for the
ISCO and spin of the final BH.

A last element to take into account is that GW emission during the
inspiral will further reduce the energy $M$ that the system has at
infinite orbital separation. This was not considered in the BKL model
(see assumption $1$ in the previous section), but we wish to include
it in our extension of their formulation. It affects the denominator
of Eq.\,(\ref{eq:step2}) and may be taken into account at first order
in the symmetric mass ratio
$\nu=M_\text{BH}M_\text{NS}/(M_\text{BH}+M_\text{NS})^2$ by
subtracting to $M$ the additional term, e.g.~\cite{Barausse2012b},
\begin{align}
  E_\text{rad}=M[1-e(\bar{r}_\text{ISCO,i},a_\text{i})]\nu\,,
\end{align}
so that 
\begin{align}
\label{eq:model}
a_\text{f} = \frac{a_\text{i}M_\text{BH}^2 +
  l_z(\bar{r}_\text{ISCO,f},a_\text{f})M_\text{BH}(M_\text{NS}-M_\text{b,torus})}{[M\left\{1-[1-e(\bar{r}_\text{ISCO,i},a_\text{i})]\nu\right\}-e(\bar{r}_\text{ISCO,f},a_\text{f})M_\text{b,torus}]^2}\,.
\end{align}
This final, closed expression for the final spin parameter
$a_\text{f}$ may be solved numerically with root-finding techniques to
determine the spin parameter of the BH remnant of BH-NS mergers, and
its denominator automatically provides a prediction for the final mass
of the remnant itself. In other words, once $a_\text{f}$ is
calculated, the mass of the BH remnant $M_\text{f}$ automatically
follows as
\begin{align}
\label{eq:model-Mf}
M_\text{f} &=
M\left\{1-[1-e(\bar{r}_\text{ISCO,i},a_\text{i})]\nu\right\}-e(\bar{r}_\text{ISCO,f},a_\text{f})M_\text{b,torus}\,.
\end{align}

Notice that, in principle, Eq.\,(\ref{eq:model}) may be generalized to
account for additional energy losses and for nonideal angular momentum
accretion. In the former case, it is sufficient to subtract extra
terms on the right-hand side of Eq.\,(\ref{eq:model-Mf}) and, hence,
in the denominator of Eq.\,(\ref{eq:model}). Nonideal angular momentum
accretion, which is particularly relevant for disruptive BH-NS
mergers, could instead be modeled by inserting an angular momentum
accretion efficiency factor in front of the $l_z$ appearing in
Eq.\,(\ref{eq:model}). For the time being, we keep
Eq.\,(\ref{eq:model}) as it is, knowing that it may be improved as the
nuances in the physics of BH-NS mergers become clearer.

%TTTTTTTTTTTTTTTTTTTTTTTTTTTTTTTTTTTTTTTTTTTTTTTTTTTTTTTTTTTTTTTTTTTTT
\begin{table}[!t]
  \caption{\label{tab:tests1} Tests against numerical-relativity results. Each row is a testcase numbered by the index in the first column. The remaining columns provide the reference in which the numerical-relativity simulation for the binary was presented, information about the NS EOS, the NS compactness $C$, the binary mass ratio $Q$, the initial BH spin parameter $a_\text{i}$, and the final BH spin parameter given by the numerical-relativity simulation, $a_\text{f}^\text{NR}$, by the BKL approach, $a_\text{f}^\text{BKL}$, by Eq.\,(\ref{eq:model}), $a_\text{f,1}$, and by the final formulation of our model given in Eqs.\,(\ref{eq:model-imp}) and (\ref{eq:fbridge}), $a_\text{f}$. The NSs in the initial data of all simulations are spinless.}
  \begin{tabular}{l@{\hspace{0.25cm}}c@{\hspace{0.25cm}}c@{\hspace{0.25cm}}c@{\hspace{0.25cm}}c@{\hspace{0.25cm}}d{2.2}@{\hspace{0.25cm}}c@{\hspace{0.25cm}}c@{\hspace{0.25cm}}c@{\hspace{0.25cm}}c@{\hspace{0.1cm}}}
    \toprule[1.pt]
    \toprule[1.pt]
    \addlinespace[0.3em]
    & Ref. & EOS & $C$ & $Q$ & \multicolumn{1}{c}{$a_\text{i}$} & $a_\text{f}^\text{NR}$ & $a_\text{f}^\text{BKL}$ & $a_\text{f,1}$ & $a_\text{f}$\\
    \addlinespace[0.2em]
    \midrule[1.pt]
    \addlinespace[0.2em]
    1 & \cite{Etienne:2008re} & $\Gamma=2$ & $0.145$ & $3$ & -0.5 & $0.33$ & $0.31$ & $0.31$ & $0.33$\\
    2 & \cite{Etienne:2008re} & $\Gamma=2$ & $0.145$ & $3$ & 0.75 & $0.88$ & $0.85$ & $0.86$ & $0.87$\\
    3 & \cite{Etienne:2008re} & $\Gamma=2$ & $0.145$ & $3$ & 0    & $0.56$ & $0.54$ & $0.53$ & $0.55$\\
    4 & \cite{Etienne:2008re} & $\Gamma=2$ & $0.145$ & $5$ & 0    & $0.42$ & $0.42$ & $0.42$ & $0.42$\\
    \midrule
    5 & \cite{Shibata2012err} & $\Gamma=2$ & $0.145$ & $2$   & 0 & $0.68$ & $0.61$ & $0.61$ & $0.64$\\
    6 & \cite{Shibata2012err} & $\Gamma=2$ & $0.145$ & $3$   & 0 & $0.56$ & $0.54$ & $0.53$ & $0.55$\\
    7 & \cite{Shibata2012err} & $\Gamma=2$ & $0.145$ & $4$   & 0 & $0.48$ & $0.47$ & $0.47$ & $0.47$\\
    8 & \cite{Shibata2012err} & $\Gamma=2$ & $0.145$ & $5$   & 0 & $0.42$ & $0.42$ & $0.42$ & $0.42$\\
    9 & \cite{Shibata2012err} & $\Gamma=2$ & $0.160$ & $2$   & 0 & $0.68$ & $0.61$ & $0.62$ & $0.65$\\
    10 & \cite{Shibata2012err} & $\Gamma=2$ & $0.160$ & $3$   & 0 & $0.55$ & $0.54$ & $0.54$ & $0.55$\\
    11 & \cite{Shibata2012err} & $\Gamma=2$ & $0.178$ & $2$   & 0 & $0.67$ & $0.61$ & $0.62$ & $0.66$\\
    12 & \cite{Shibata2012err} & $\Gamma=2$ & $0.178$ & $3$   & 0 & $0.55$ & $0.54$ & $0.54$ & $0.56$\\
    \midrule
    13 & \cite{Foucart2010} & $\Gamma=2$ & $0.144$ & $3$ & 0   & $0.56$ & $0.54$ & $0.53$ & $0.54$\\
    14 & \cite{Foucart2010} & $\Gamma=2$ & $0.144$ & $3$ & 0.5 & $0.77$ & $0.75$ & $0.75$ & $0.76$\\
    15 & \cite{Foucart2010} & $\Gamma=2$ & $0.144$ & $3$ & 0.9 & $0.93$ & $0.90$ & $0.93$ & $0.93$\\
    \midrule
    16 & \cite{Foucart2011} & $\Gamma=2$ & $0.144$ & $7$ & 0.5 & $0.67$ & $0.67$ & $0.68$ & $0.68$\\
    17 & \cite{Foucart2011} & $\Gamma=2$ & $0.144$ & $7$ & 0.7 & $0.80$ & $0.80$ & $0.81$ & $0.81$\\
    18 & \cite{Foucart2011} & $\Gamma=2$ & $0.144$ & $7$ & 0.9 & $0.92$ & $0.91$ & $0.93$ & $0.93$\\
    19 & \cite{Foucart2011} & $\Gamma=2$ & $0.144$ & $5$ & 0.5 & $0.71$ & $0.71$ & $0.71$ & $0.71$\\
    \midrule
    20 & \cite{Kyutoku2011err} & 2H-135   & $0.131$ & $2$ & 0 & $0.64$ & $0.61$ & $0.59$ & $0.63$\\
    21 & \cite{Kyutoku2011err} & H-135    & $0.162$ & $2$ & 0 & $0.67$ & $0.61$ & $0.61$ & $0.66$\\
    22 & \cite{Kyutoku2011err} & HB-135   & $0.172$ & $2$ & 0 & $0.67$ & $0.61$ & $0.62$ & $0.66$\\
    23 & \cite{Kyutoku2011err} & HBs-135  & $0.172$ & $2$ & 0 & $0.67$ & $0.61$ & $0.62$ & $0.66$\\
    24 & \cite{Kyutoku2011err} & HBss-135 & $0.174$ & $2$ & 0 & $0.67$ & $0.61$ & $0.62$ & $0.66$\\
    25 & \cite{Kyutoku2011err} & B-135    & $0.182$ & $2$ & 0 & $0.67$ & $0.61$ & $0.62$ & $0.67$\\
    26 & \cite{Kyutoku2011err} & Bs-135   & $0.185$ & $2$ & 0 & $0.66$ & $0.61$ & $0.62$ & $0.67$\\
    27 & \cite{Kyutoku2011err} & Bss-135  & $0.194$ & $2$ & 0 & $0.65$ & $0.61$ & $0.62$ & $0.67$\\
    28 & \cite{Kyutoku2011err} & 2H-135   & $0.131$ & $3$ & 0 & $0.52$ & $0.54$ & $0.51$ & $0.52$\\
    29 & \cite{Kyutoku2011err} & H-135    & $0.162$ & $3$ & 0 & $0.56$ & $0.54$ & $0.54$ & $0.56$\\
    30 & \cite{Kyutoku2011err} & HB-135   & $0.172$ & $3$ & 0 & $0.56$ & $0.54$ & $0.54$ & $0.56$\\
    31 & \cite{Kyutoku2011err} & B-135    & $0.182$ & $3$ & 0 & $0.55$ & $0.54$ & $0.54$ & $0.56$\\
    32 & \cite{Kyutoku2011err} & 2H-12    & $0.118$ & $2$ & 0 & $0.62$ & $0.61$ & $0.58$ & $0.62$\\
    33 & \cite{Kyutoku2011err} & H-12     & $0.145$ & $2$ & 0 & $0.66$ & $0.61$ & $0.60$ & $0.64$\\
    34 & \cite{Kyutoku2011err} & HB-12    & $0.153$ & $2$ & 0 & $0.66$ & $0.61$ & $0.61$ & $0.65$\\
    35 & \cite{Kyutoku2011err} & B-12     & $0.161$ & $2$ & 0 & $0.67$ & $0.61$ & $0.61$ & $0.66$\\
    36 & \cite{Kyutoku2011err} & HB-12    & $0.153$ & $3$ & 0 & $0.55$ & $0.54$ & $0.54$ & $0.55$\\
    37 & \cite{Kyutoku2011err} & B-12     & $0.161$ & $3$ & 0 & $0.56$ & $0.54$ & $0.54$ & $0.56$\\
    \bottomrule[1.pt]
    \bottomrule[1.pt]
  \end{tabular}
\end{table}
%TTTTTTTTTTTTTTTTTTTTTTTTTTTTTTTTTTTTTTTTTTTTTTTTTTTTTTTTTTTTTTTTTTTTT

%TTTTTTTTTTTTTTTTTTTTTTTTTTTTTTTTTTTTTTTTTTTTTTTTTTTTTTTTTTTTTTTTTTTTT
\begin{table}[!t]
  \caption{\label{tab:tests2} Same as Table \ref{tab:tests1}.}
\begin{tabular}{l@{\hspace{0.25cm}}c@{\hspace{0.25cm}}c@{\hspace{0.25cm}}c@{\hspace{0.25cm}}c@{\hspace{0.25cm}}d{2.2}@{\hspace{0.25cm}}c@{\hspace{0.25cm}}c@{\hspace{0.25cm}}c@{\hspace{0.25cm}}c@{\hspace{0.1cm}}}
    \toprule[1.pt]
    \toprule[1.pt]
    \addlinespace[0.3em]
    & Ref. & EOS & $C$ & $Q$ & \multicolumn{1}{c}{$a_\text{i}$} & $a_\text{f}^\text{NR}$ & $a_\text{f}^\text{BKL}$ & $a_\text{f,1}$ & $a_\text{f}$\\
    \addlinespace[0.2em]
    \midrule[1.pt]
    \addlinespace[0.2em]
    38 & \cite{Kyutoku2011} & 2H-135   & $0.131$ & $2$ & 0.75 & $0.87$ & $0.84$ & $0.86$ & $0.89$\\
    39 & \cite{Kyutoku2011} & 1.5H-135 & $0.146$ & $2$ & 0.75 & $0.89$ & $0.84$ & $0.86$ & $0.89$\\
    40 & \cite{Kyutoku2011} & H-135    & $0.162$ & $2$ & 0.75 & $0.91$ & $0.84$ & $0.86$ & $0.90$\\
    41 & \cite{Kyutoku2011} & HB-135   & $0.172$ & $2$ & 0.75 & $0.91$ & $0.84$ & $0.86$ & $0.90$\\
    42 & \cite{Kyutoku2011} & B-135    & $0.182$ & $2$ & 0.75 & $0.91$ & $0.84$ & $0.86$ & $0.90$\\
    43 & \cite{Kyutoku2011} & 2H-135   & $0.131$ & $2$ & 0.5  & $0.81$ & $0.77$ & $0.77$ & $0.80$\\
    44 & \cite{Kyutoku2011} & 1.5H-135 & $0.146$ & $2$ & 0.5  & $0.82$ & $0.77$ & $0.77$ & $0.81$\\
    45 & \cite{Kyutoku2011} & H-135    & $0.162$ & $2$ & 0.5  & $0.82$ & $0.77$ & $0.78$ & $0.82$\\
    46 & \cite{Kyutoku2011} & HB-135   & $0.172$ & $2$ & 0.5  & $0.83$ & $0.77$ & $0.78$ & $0.82$\\
    47 & \cite{Kyutoku2011} & B-135    & $0.182$ & $2$ & 0.5  & $0.83$ & $0.77$ & $0.78$ & $0.82$\\
    48 & \cite{Kyutoku2011} & 2H-135   & $0.131$ & $2$ & -0.5 & $0.48$ & $0.44$ & $0.42$ & $0.46$\\
    49 & \cite{Kyutoku2011} & H-135    & $0.162$ & $2$ & -0.5 & $0.51$ & $0.44$ & $0.45$ & $0.50$\\
    50 & \cite{Kyutoku2011} & HB-135   & $0.172$ & $2$ & -0.5 & $0.50$ & $0.44$ & $0.45$ & $0.50$\\
    51 & \cite{Kyutoku2011} & B-135    & $0.182$ & $2$ & -0.5 & $0.49$ & $0.44$ & $0.45$ & $0.51$\\
    52 & \cite{Kyutoku2011} & 2H-135   & $0.131$ & $3$ & 0.75 & $0.86$ & $0.85$ & $0.86$ & $0.87$\\
    53 & \cite{Kyutoku2011} & 1.5H-135 & $0.146$ & $3$ & 0.75 & $0.86$ & $0.85$ & $0.86$ & $0.87$\\
    54 & \cite{Kyutoku2011} & H-135    & $0.162$ & $3$ & 0.75 & $0.85$ & $0.85$ & $0.86$ & $0.87$\\
    55 & \cite{Kyutoku2011} & HB-135   & $0.172$ & $3$ & 0.75 & $0.87$ & $0.85$ & $0.86$ & $0.88$\\
    56 & \cite{Kyutoku2011} & B-135    & $0.182$ & $3$ & 0.75 & $0.86$ & $0.85$ & $0.87$ & $0.88$\\
    57 & \cite{Kyutoku2011} & 2H-135   & $0.131$ & $3$ & 0.5  & $0.74$ & $0.75$ & $0.74$ & $0.75$\\
    58 & \cite{Kyutoku2011} & 1.5H-135 & $0.146$ & $3$ & 0.5  & $0.75$ & $0.75$ & $0.75$ & $0.76$\\
    59 & \cite{Kyutoku2011} & H-135    & $0.162$ & $3$ & 0.5  & $0.76$ & $0.75$ & $0.75$ & $0.77$\\
    60 & \cite{Kyutoku2011} & HB-135   & $0.172$ & $3$ & 0.5  & $0.77$ & $0.75$ & $0.76$ & $0.77$\\
    61 & \cite{Kyutoku2011} & B-135    & $0.182$ & $3$ & 0.5  & $0.77$ & $0.75$ & $0.76$ & $0.78$\\
    62 & \cite{Kyutoku2011} & HB-135   & $0.172$ & $3$ & -0.5 & $0.32$ & $0.31$ & $0.31$ & $0.34$\\
    63 & \cite{Kyutoku2011} & 2H-135   & $0.131$ & $4$ & 0.75 & $0.84$ & $0.84$ & $0.85$ & $0.85$\\
    64 & \cite{Kyutoku2011} & H-135    & $0.162$ & $4$ & 0.75 & $0.84$ & $0.84$ & $0.86$ & $0.86$\\
    65 & \cite{Kyutoku2011} & HB-135   & $0.172$ & $4$ & 0.75 & $0.85$ & $0.84$ & $0.86$ & $0.86$\\
    66 & \cite{Kyutoku2011} & B-135    & $0.182$ & $4$ & 0.75 & $0.85$ & $0.84$ & $0.86$ & $0.86$\\
    67 & \cite{Kyutoku2011} & 2H-135   & $0.131$ & $4$ & 0.5  & $0.70$ & $0.73$ & $0.71$ & $0.71$\\
    68 & \cite{Kyutoku2011} & H-135    & $0.162$ & $4$ & 0.5  & $0.73$ & $0.73$ & $0.73$ & $0.73$\\
    69 & \cite{Kyutoku2011} & HB-135   & $0.172$ & $4$ & 0.5  & $0.74$ & $0.73$ & $0.74$ & $0.74$\\
    70 & \cite{Kyutoku2011} & B-135    & $0.182$ & $4$ & 0.5  & $0.74$ & $0.73$ & $0.74$ & $0.74$\\
    71 & \cite{Kyutoku2011} & 2H-135   & $0.131$ & $5$ & 0.75 & $0.82$ & $0.84$ & $0.84$ & $0.84$\\
    72 & \cite{Kyutoku2011} & H-135    & $0.162$ & $5$ & 0.75 & $0.84$ & $0.84$ & $0.85$ & $0.85$\\
    73 & \cite{Kyutoku2011} & HB-135   & $0.172$ & $5$ & 0.75 & $0.84$ & $0.84$ & $0.85$ & $0.85$\\
    74 & \cite{Kyutoku2011} & B-135    & $0.182$ & $5$ & 0.75 & $0.85$ & $0.84$ & $0.86$ & $0.86$\\
    \bottomrule[1.pt]
    \bottomrule[1.pt]
  \end{tabular}
\end{table}
%TTTTTTTTTTTTTTTTTTTTTTTTTTTTTTTTTTTTTTTTTTTTTTTTTTTTTTTTTTTTTTTTTTTTT

In Tables \ref{tab:tests1} and \ref{tab:tests2} we compare the
predictions of Eq.\,(\ref{eq:model}) and Eq.\,(\ref{eq:BKL}) to the
results obtained within full general relativity
in~\cite{Etienne:2008re, Foucart2010, Foucart2011, Kyutoku2011,
  Kyutoku2011err, Shibata2012err}, which, along
with~\cite{Etienne2012, Lackey2012, Etienne2012b, Deaton2013,
  Foucart2013a, Kyutoku2013, Foucart2013b, Lovelace2013,
  Paschalidis2013b}, represent the state of the art of
numerical-relativity simulations of BH-NS mergers. Each row of the
tables refers to a specific BH-NS binary coalescence. The columns
provide a dummy index which numbers the test cases, the reference in
which the numerical-relativity results for that binary were presented,
information about the NS EOS, the NS compactness
$C=M_\text{NS}/R_\text{NS}$, the binary mass ratio
$Q=M_\text{BH}/M_\text{NS}$, the initial BH spin parameter
$a_\text{i}$, the numerical-relativity result for the final BH spin
parameter $a_\text{f}^\text{NR}$, the final BH spin parameter
$a_\text{f}^\text{BKL}$ predicted by the BKL formula in
Eq.\,(\ref{eq:BKL}), the final BH spin parameter $a_\text{f,1}$
yielded by Eq.\,(\ref{eq:model}), and the final BH spin parameter
$a_\text{f}$ predicted by Eq.\,(\ref{eq:model-imp}), which contains
improvements over Eq.\,(\ref{eq:model}) and will be discussed
later. As far as the NS EOS is concerned, the first $19$ comparisons
reported in Table \ref{tab:tests1} refer to binaries in which the
nonthermal behavior of the NS matter\footnote{Here and
  in~\cite{Pannarale2010, Foucart2012} thermal contributions are
  neglected. These are more relevant, in the merger and postmerger
  dynamics, when the NS is tidally disrupted.} is governed, at
microphysical level, by a polytropic EOS with polytropic exponent
$\Gamma=2$. In the last $18$ simulations reported in Table
\ref{tab:tests1} and in all the ones reported in Table
\ref{tab:tests2}, on the other hand, a two-piecewise polytropic EOS
was used, and the notation in the tables follows the one
of~\cite{Kyutoku2010, Kyutoku2011}: the first half of the label
indicates the stiffness of the EOS, with 2H being the stiffest,
whereas the second half refers to the NS ADM mass at isolation
(e.g. $135$ stands for $1.35M_\odot$). In this first round of tests,
we used the values of $M_\text{b,torus}$ found with the
numerical-relativity simulations and reported in the papers. To make
the whole model numerical relativity independent and quick to use, we
shall later adopt the method recently reported in~\cite{Foucart2012}
for determining $M_\text{b,torus}$, and we will show that this does
not spoil the agreement between the predictions of our model and the
numerical-relativity data. It is evident that the difference between
$a_\text{f,1}$ and $a_\text{f}^\text{NR}$ increases as the mass ratio
$Q$ of the system decreases, or, equivalently, as the symmetric mass
ratio $\nu$ increases. Given that the final spin parameter results
obtained with numerical-relativity simulations have an absolute error
$\Delta a_\text{f}^\text{NR}$ of $0.01$~\cite{Shibata:2009cn} and that
the error of the BKL approach was evaluated to be $\lesssim 0.02$ in
\cite{Barausse:2009uz}, we conclude that the method established by
Eq.\,(\ref{eq:model}) works well for BH-NS systems with symmetric mass
ratios up to $\nu=0.16$, i.e. for $Q\geq 4$, whereas it almost
systematically reaches or exceeds the $0.03$ threshold of marginal
agreement when $Q\leq 3$. We must, thus, improve Eq.\,(\ref{eq:model})
to handle BH-NS systems with $\nu > 0.16$.

As $\nu$ increases, the method fails for two reasons. First, the fifth
assumption in Sec.\,\ref{sec:BKL} breaks down as $\nu\rightarrow 0.25$
(or $Q\rightarrow 1$). This is intrinsic to the BKL method which
inspired Eq.\,(\ref{eq:model}). Second, and generally speaking, in
systems with such low mass BHs the tidal fields tend to tear apart the
NS completely, as opposed to binaries with higher mass BHs, in which
the outer layers of the NS are mainly stripped off. In the former
scenario, the binding energy of the star is liberated and the NS
matter accretes onto the BH as a collection of particles with total
rest mass $M_\text{b,NS}-M_\text{b,torus}$, where $M_\text{b,NS}$ is
the total rest mass of the NS, whereas in the latter scenario the NS
core plunges into the BH without undergoing complete tidal
disruption. We will make the simplifying assumption that in systems
with $\nu=2/9$ ($Q=2$), the NS undergoes complete tidal disruption,
while it does not in systems with $\nu\leq 0.16$ ($Q\geq 4$). When
complete tidal disruption is achieved, the NS should not be treated as
a body with mass $M_\text{NS}$, but as a set of particles with total
rest mass $M_\text{b,NS}$, a subset of which accretes onto the BH and
has total mass $M_\text{b,NS}-M_\text{b,torus}$. We thus propose to
describe $Q=2$ systems, in which tidal disruption is pivotal, with
\begin{align}
\label{eq:model-Q2}
a_\text{f} = \frac{a_\text{i}M_\text{BH}^2 +
  l_z(\bar{r}_\text{ISCO,f},a_\text{f})M_\text{BH}(M_\text{b,NS}-M_\text{b,torus})}{[M\left\{1-[1-e(\bar{r}_\text{ISCO,i},a_\text{i})]\nu\right\}-e(\bar{r}_\text{ISCO,f},a_\text{f})M_\text{b,torus}]^2},
\end{align}
instead of with Eq.\,(\ref{eq:model}), and to combine the two
descriptions by writing
\begin{widetext}
\begin{align}
\label{eq:model-imp}
a_\text{f} = \frac{a_\text{i}M_\text{BH}^2 +
  l_z(\bar{r}_\text{ISCO,f},a_\text{f})M_\text{BH}\{[1-f(\nu)]M_\text{NS}+f(\nu)M_\text{b,NS}-M_\text{b,torus}\}}{[M\left\{1-[1-e(\bar{r}_\text{ISCO,i},a_\text{i})]\nu\right\}-e(\bar{r}_\text{ISCO,f},a_\text{f})M_\text{b,torus}]^2}\,,
\end{align}
\end{widetext}
where $f(\nu)$ governs the transition between the two regimes of
Eqs.\,(\ref{eq:model}) and (\ref{eq:model-Q2}). This function is
currently poorly constrained, given that state-of-the-art BH-NS
simulations with $2<Q<4$ are available in the literature only for
$Q=3$. To fix $f(\nu)$, we must impose that $f(\nu\geq 2/9)=1$ and
that $f(\nu\leq 0.16)=0$. Additionally, it is physically reasonable to
require the function to be monotonic and therefore that
\begin{eqnarray}
  \frac{df}{d\nu} \geq 0  & & ~~~~~~~ 0 \leq\nu\leq 0.25\,.\nn
\end{eqnarray}
%
%FFFFFFFFFFFFFFFFFFFFFFFFFFFFFFFFFFFFFFFFFFFFFFFFFFFFFFFFFFFFFFFFFFFFF
\begin{figure}[!b]
  \includegraphics[width=8cm]{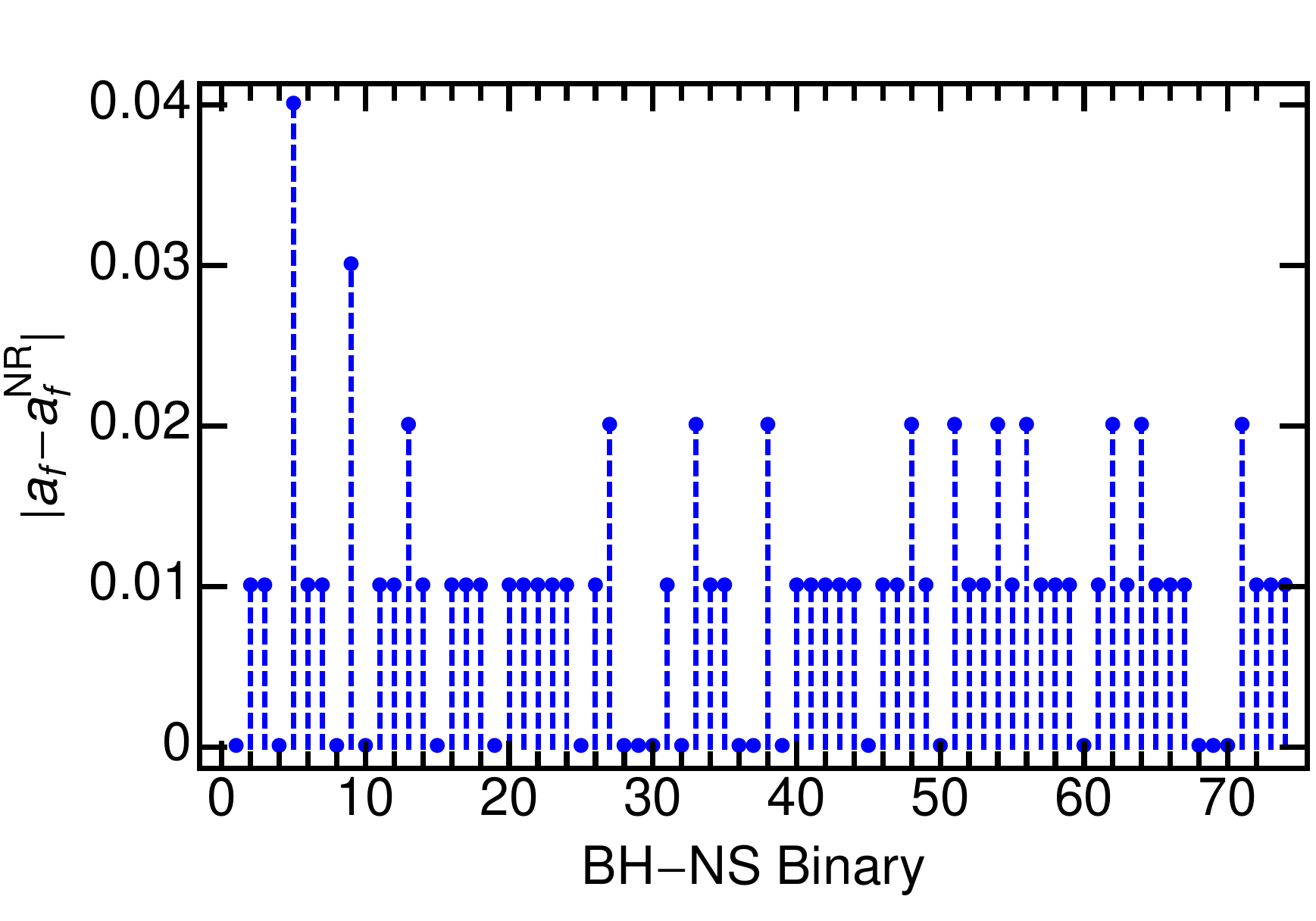}
  \caption{\label{FIG:AbsErrors} $|a_\text{f}-a_\text{f}^\text{NR}|$
    is shown for all entries in Tables \ref{tab:tests1} and
    \ref{tab:tests2}. The horizontal axis is the dummy that runs
    through both tables.}
\end{figure}
%FFFFFFFFFFFFFFFFFFFFFFFFFFFFFFFFFFFFFFFFFFFFFFFFFFFFFFFFFFFFFFFFFFFFF

\noindent We shall also require it to be $C^\infty$ and to be as
simple as possible. These elements do not determine $f(\nu)$ uniquely,
of course. All in all, we set
\begin{align}
\label{eq:fbridge}
f(\nu) = \left\{
\begin{array}{ll}
  0 & \nu \leq 0.16 \\
  \frac{1}{2}\Big[1-\cos\Big(\frac{\pi(\nu - 0.16)}{2/9-0.16}\Big)\Big] & 0.16<\nu<2/9 \\
  1 & 2/9\leq\nu\leq 0.25 \\
\end{array}
\right.\nn\\
\end{align}
in a Hann window inspired fashion. Notice that, in the limit of large
BH masses, a BH-NS system behaves as a BH binary system with the same
physical parameters, so that in Eq.\,(\ref{eq:model}) the NS
gravitational mass $M_\text{NS}$ cannot be simply dropped in favor of
its baryonic mass $M_\text{b,NS}$. Moreover, from a merely
quantitative point of view, a model with this oversimplification
performs worse when tested against numerical-relativity results.

We now compare the predictions of
Eqs.\,(\ref{eq:model-imp})-(\ref{eq:fbridge}) to the results obtained
within full general relativity. As anticipated, this is done in the
last column of Tables \ref{tab:tests1} and \ref{tab:tests2}. It is
evident that this strategy improves considerably the outcome of
Eq.\,(\ref{eq:model}) for systems with $Q=2$ and $Q=3$, and that,
overall, it improves the estimates obtained by simply applying the BKL
method to mixed binary mergers.

%FFFFFFFFFFFFFFFFFFFFFFFFFFFFFFFFFFFFFFFFFFFFFFFFFFFFFFFFFFFFFFFFFFFFF
\begin{figure}[!b]
  \includegraphics[width=8cm]{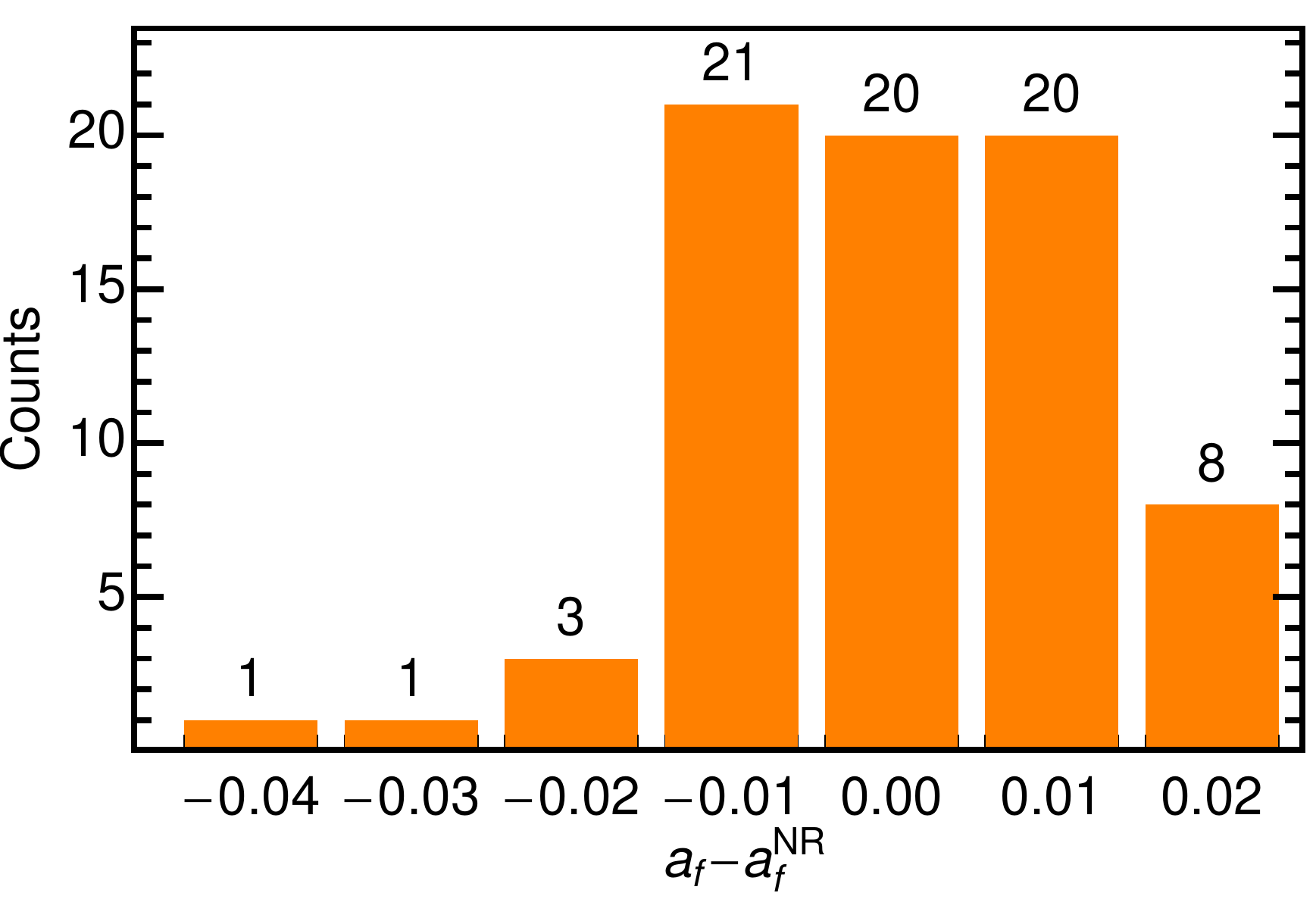}
  \caption{\label{FIG:HistoErrors} $a_\text{f}-a_\text{f}^\text{NR}$
    distribution for all entries in Tables \ref{tab:tests1} and
    \ref{tab:tests2}.}
\end{figure}
%FFFFFFFFFFFFFFFFFFFFFFFFFFFFFFFFFFFFFFFFFFFFFFFFFFFFFFFFFFFFFFFFFFFFF

Figure \ref{FIG:AbsErrors} shows the absolute value of the difference
$a_\text{f}-a_\text{f}^\text{NR}$ versus the dummy index running over
the 74 rows of Tables \ref{tab:tests1} and \ref{tab:tests2}. The graph
shows that $\max|a_\text{f}-a_\text{f}^\text{NR}|=0.04$ and that this
value is reached only in one case out of 74 total ones. This
corresponds to the $\{C=0.145,Q=2,a_\text{i}=0\}$ simulation
of~\cite{Shibata:2009cn,Shibata2012err}. An absolute error
$|a_\text{f}-a_\text{f}^\text{NR}|=0.03$ is instead obtained for the
test case $\{C=0.160,Q=2,a_\text{i}=0\}$
of~\cite{Shibata:2009cn,Shibata2012err}. We notice that both
problematic cases have $Q=2$ and this may be a sign that our model
breaks down for low BH masses.

%FFFFFFFFFFFFFFFFFFFFFFFFFFFFFFFFFFFFFFFFFFFFFFFFFFFFFFFFFFFFFFFFFFFFF
\begin{figure}[!t]
  \includegraphics[width=8cm]{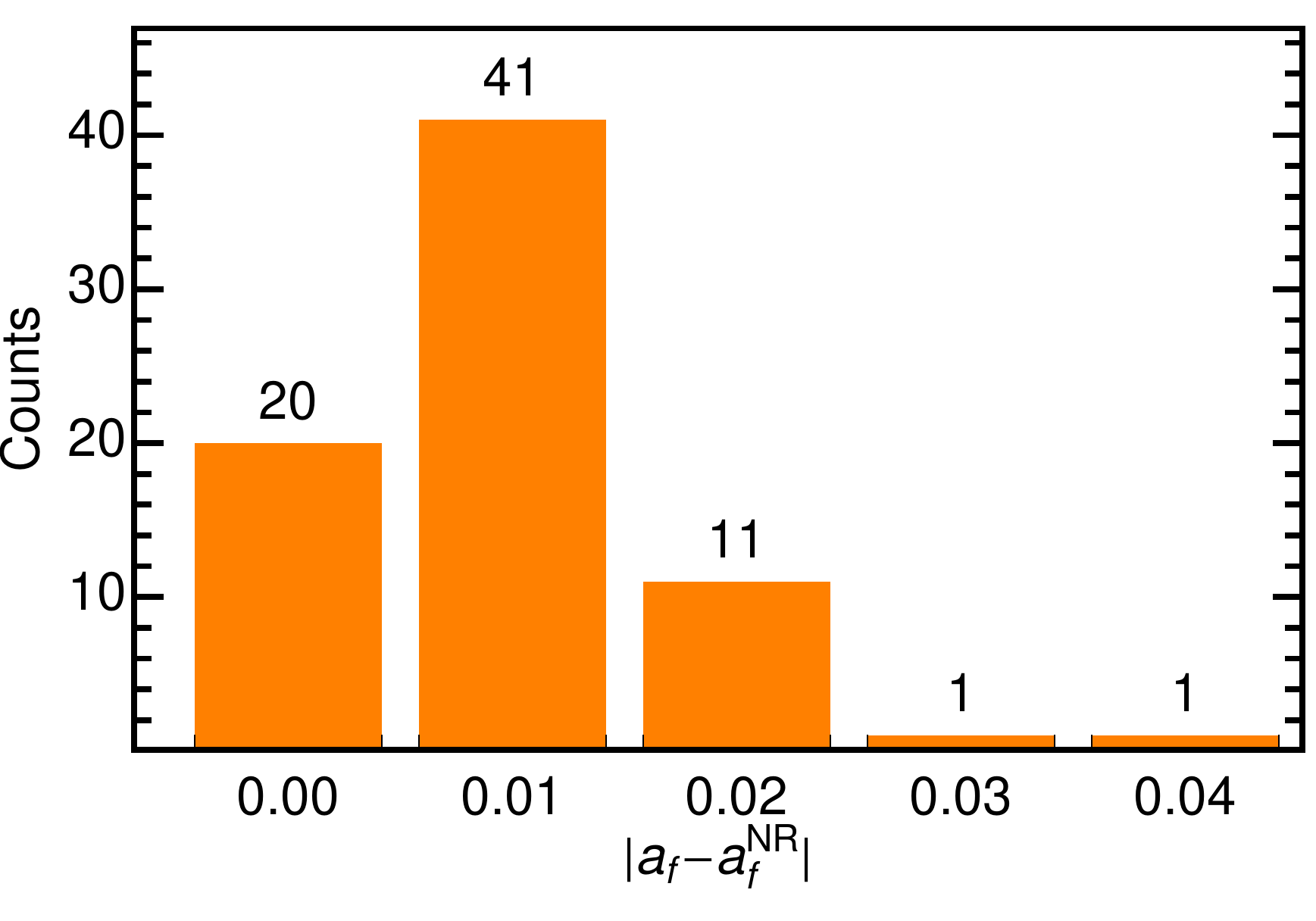}
  \caption{\label{FIG:HistoAbsErrors}
    $|a_\text{f}-a_\text{f}^\text{NR}|$ distribution for all entries
    in Tables \ref{tab:tests1} and \ref{tab:tests2}.}
\end{figure}
%FFFFFFFFFFFFFFFFFFFFFFFFFFFFFFFFFFFFFFFFFFFFFFFFFFFFFFFFFFFFFFFFFFFFF

In Fig.\,\ref{FIG:HistoErrors}, we show the distribution of the
difference $a_\text{f}-a_\text{f}^\text{NR}$ that follows from the
results reported in Tables \ref{tab:tests1} and \ref{tab:tests2}. As
is evident, the errors are concentrated in the interval $-0.01\leq
a_\text{f}-a_\text{f}^\text{NR}\leq 0.01$, where $61$ of the $74$
tests fall. Moreover, about a quarter of the final spin values are
reproduced exactly. The sum of all the differences
$a_\text{f}-a_\text{f}^\text{NR}$ yields $0.02$, so that
$\sum_{n=1}^{74}(a_\text{f}-a_\text{f}^\text{NR})_n/74=0.00$\footnote{This
  rounding up is justified by the fact that numerical-relativity
  results for the final spin parameter have an error $\Delta
  a_\text{f}^\text{NR}\sim 0.01$.}. In Fig.\,\ref{FIG:HistoAbsErrors},
we consider the distribution of the absolute difference
$|a_\text{f}-a_\text{f}^\text{NR}|$ and show that it rapidly drops
after $0.02$. Given that the error on the $a_\text{f}^\text{NR}$'s is
$\Delta a_\text{f}^\text{NR}=0.01$, $61$ numerical-relativity results
out of $74$, i.e. more than $80$\% of the cases, are reproduced within
the numerical-relativity error.

To determine the absolute error on our predictions, we begin by
observing that: (1) our model is built against numerical-relativity
data with an absolute error $\Delta a_\text{f}^\text{NR}=0.01$; (2) it
is based on the BKL approach, for which the average error found in
\cite{Barausse:2009uz} is $\sim 0.02$; and (3) the average error
yielded by our comparisons against numerical-relativity data is
$\langle
|a_\text{f}-a_\text{f}^\text{NR}|\rangle=\sum_{n=1}^{74}|a_\text{f}-a_\text{f}^\text{NR}|_n/74=0.01$. Further,
the same average error is obtained if the average $\langle
|a_\text{f}-a_\text{f}^\text{NR}|\rangle$ is marginalized to a given
mass ratio $Q$ or initial BH spin parameter value $a_\text{i}$, among
the ones available in Tables \ref{tab:tests1} and
\ref{tab:tests2}. This is shown in the second column of Table
\ref{tab:qFix-aFix}. Additionally, the fourth column of the same table
shows that the marginalized averages $\langle
a_\text{f}-a_\text{f}^\text{NR}\rangle$ are such that their absolute
value is $0.01$ at the most. All in all, if we take an error $\Delta
a_\text{f}=0.01$ on our predictions, these are found to be compatible
with the numerical-relativity results in $72$ test cases out of $74$,
i.e. about $97$\%.

%TTTTTTTTTTTTTTTTTTTTTTTTTTTTTTTTTTTTTTTTTTTTTTTTTTTTTTTTTTTTTTTTTTTTT
\begin{table}[!t]
  \caption{\label{tab:qFix-aFix} Average $|a_\text{f}-a_\text{f}^\text{NR}|$ and $a_\text{f}-a_\text{f}^\text{NR}$ for a given physical parameter, indicated in the first column. The second and fourth columns refer to predictions for $a_\text{f}$ obtained by substituting for $M_\text{b,torus}$ in Eq.\,(\ref{eq:model-imp}) the results found in the numerical-relativity simulations reported in~\cite{Etienne:2008re, Foucart2010, Foucart2011, Kyutoku2010, Kyutoku2011err, Shibata2012err}, whereas the third and fifth columns refer to predictions for $a_\text{f}$ obtained by using the model described in~\cite{Foucart2012} to calculate the $M_\text{b,torus}$'s.}
  \begin{tabular}{@{\hspace{0.45cm}}c@{\hspace{0.7cm}}c@{\hspace{0.7cm}}c@{\hspace{0.7cm}}r@{\hspace{0.7cm}}r@{\hspace{0.45cm}}}
    \toprule[1.pt]
    \toprule[1.pt]
    \addlinespace[0.3em]
    Fix parameter & \multicolumn{2}{c}{$\langle |a_\text{f}-a_\text{f}^\text{NR}|\rangle$~~~~~~~} & \multicolumn{2}{c}{$\langle a_\text{f}-a_\text{f}^\text{NR}\rangle$}\\
    \addlinespace[0.2em]
    \midrule[1.pt]
    \addlinespace[0.2em]
    $Q=2$ & $0.01$ & $0.01$ & \hspace{0.3cm} $-0.01$ & $-0.01$ \\
    $Q=3$ & $0.01$ & $0.01$ & $0.00$  & $0.01$ \\
    $Q=4$ & $0.01$ & $0.01$ & $0.01$  & $0.01$ \\
    $Q=5$ & $0.01$ & $0.01$ & $0.01$  & $0.01$ \\
    $Q=7$ & $0.01$ & $0.01$ & $0.01$  & $0.01$ \\
    \midrule[0.5pt]
    $a_\text{i}=-0.5$ & $0.01$ & $0.01$ & $0.00$ & $0.00$ \\
    $a_\text{i}=0$    & $0.01$ & $0.01$ & $-0.01$ & $-0.01$ \\
    $a_\text{i}=0.5$  & $0.01$ & $0.01$ & $0.00$  & $0.00$  \\
    $a_\text{i}=0.7$  & $0.01$ & $0.01$ & $0.01$  & $0.01$  \\
    $a_\text{i}=0.75$ & $0.01$ & $0.01$ & $0.01$  & $0.01$  \\
    $a_\text{i}=0.9$  & $0.01$ & $0.01$ & $0.01$  & $0.01$  \\
    \bottomrule[1.pt]
    \bottomrule[1.pt]
  \end{tabular}
\end{table}
%TTTTTTTTTTTTTTTTTTTTTTTTTTTTTTTTTTTTTTTTTTTTTTTTTTTTTTTTTTTTTTTTTTTTT

%FFFFFFFFFFFFFFFFFFFFFFFFFFFFFFFFFFFFFFFFFFFFFFFFFFFFFFFFFFFFFFFFFFFFF
\begin{figure}[!b]
  \includegraphics[width=8cm]{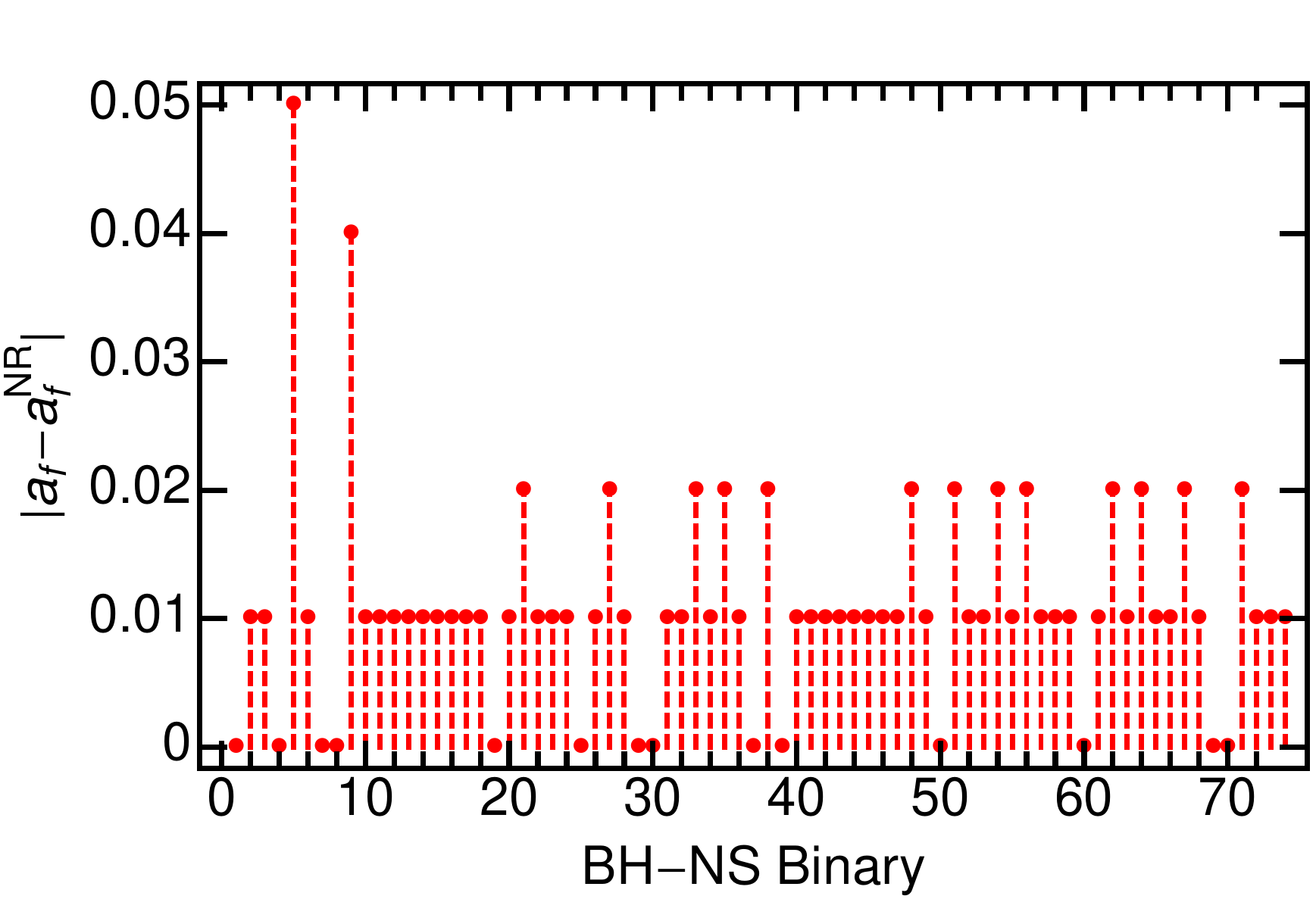}
  \caption{\label{FIG:AbsErrorsToy} Same as Fig.\,\ref{FIG:AbsErrors},
    but using the predictions of~\cite{Foucart2012} for
    $M_\text{b,torus}$ in Eq.\,(\ref{eq:model-imp}).}
\end{figure}
%FFFFFFFFFFFFFFFFFFFFFFFFFFFFFFFFFFFFFFFFFFFFFFFFFFFFFFFFFFFFFFFFFFFFF

So far, when comparing the predictions of
Eqs.\,(\ref{eq:model-imp})-(\ref{eq:fbridge}) to the BH-NS merger
results available in the literature, we exploited the
numerical-relativity prediction for $M_\text{b,torus}$. This allowed
us to test and validate Eqs.\,(\ref{eq:model-imp})-(\ref{eq:fbridge})
and to determine the average error $\Delta a_\text{f}^\text{NR}\simeq
0.01$. If we wish to apply such a method to a large number and variety
of BH-NS binaries, we must consider another way of obtaining
$M_\text{b,torus}$. As mentioned previously, we choose to use the
simple two-parameter model, fitted to existing numerical results,
recently reported by Foucart in~\cite{Foucart2012}. This provides an
estimate for $M_\text{b,torus}$, given a binary mass ratio, an initial
BH spin parameter, and a NS compactness. In
Fig.\,\ref{FIG:AbsErrorsToy} we show the absolute values of the
difference $a_\text{f}-a_\text{f}^\text{NR}$ obtained when using the
approach of~\cite{Foucart2012} to calculate $M_\text{b,torus}$; this
figure must be compared to Fig.\,\ref{FIG:AbsErrors}. We find that the
two problematic test cases, i.e.~ones with
$|a_\text{f}-a_\text{f}^\text{NR}|>0.02$, are the same ones
encountered previously, that is, cases $5$ and $9$, and that this time
they yield $|a_\text{f}-a_\text{f}^\text{NR}|$ equal to $0.05$ and
$0.04$, respectively.

\begin{figure}[!b]
  \includegraphics[width=8cm]{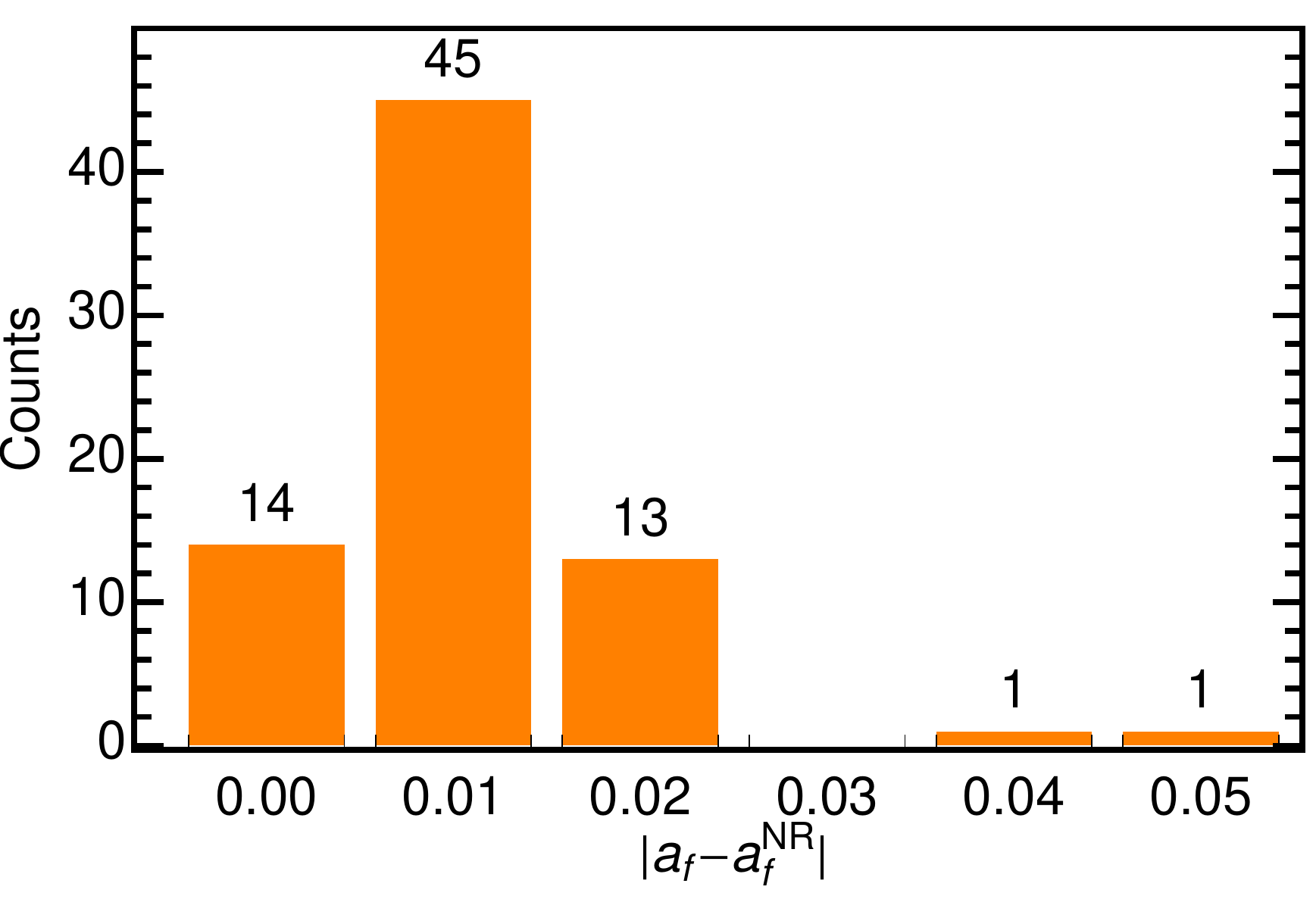}
  \caption{\label{FIG:HistoAbsErrorsToy}
    $|a_\text{f}-a_\text{f}^\text{NR}|$ distribution obtained when
    using the predictions of~\cite{Foucart2012} for $M_\text{b,torus}$
    in Eq.\,(\ref{eq:model-imp}).}
\end{figure}

As far as the distribution of the differences
$a_\text{f}-a_\text{f}^\text{NR}$ is concerned, the sum over all
differences $a_\text{f}-a_\text{f}^\text{NR}$ yields $0.04$ (as
opposed to $0.02$). The averages
$\sum_{n=1}^{74}(a_\text{f}-a_\text{f}^\text{NR})_n/74$ and
$\sum_{n=1}^{74}|a_\text{f}-a_\text{f}^\text{NR}|_n/74$ are $0.00$ and
$0.01$, respectively. The averages $\langle
|a_\text{f}-a_\text{f}^\text{NR}|\rangle$ and $\langle
a_\text{f}-a_\text{f}^\text{NR}\rangle$ marginalized to a given binary
mass ratio or an initial BH spin parameter are reported in the third
and fifth columns of Table \ref{tab:qFix-aFix}, respectively. Their
behavior does not vary significantly from the analysis reported
previously. The $|a_\text{f}-a_\text{f}^\text{NR}|$ distribution
obtained combining Eqs.\,(\ref{eq:model-imp})-(\ref{eq:fbridge}) with
the model of~\cite{Foucart2012} is shown in
Fig.\,\ref{FIG:HistoAbsErrorsToy} and should be compared to the one in
Fig.\,\ref{FIG:HistoAbsErrors}. The distribution is once again peaked
at $0.01$, and it falls off above $0.02$. Recalling that $\Delta
a_\text{f}^\text{NR}=0.01$, an agreement within the
numerical-relativity error is found in $59$ (as opposed to $61$) cases
out of $74$. If, moreover, we take an error $\Delta a_\text{f}=0.01$
on our predictions, we observe once again that $72$ of them out of
$74$ are compatible with the numerical-relativity results.

%FFFFFFFFFFFFFFFFFFFFFFFFFFFFFFFFFFFFFFFFFFFFFFFFFFFFFFFFFFFFFFFFFFFFF
%FFFFFFFFFFFFFFFFFFFFFFFFFFFFFFFFFFFFFFFFFFFFFFFFFFFFFFFFFFFFFFFFFFFFF

In conclusion, the tests and analyses performed show that the model
formulated in Eqs.\,(\ref{eq:model-imp})-(\ref{eq:fbridge}) is
robust. Using our tests against numerical-relativity results, we
argued that $\Delta a_\text{f}=0.01$. A more conservative statement is
that the error on our prediction for the final spin parameter
$a_\text{f}$ is $\Delta a_\text{f}\lesssim 0.02$. This allows us to
include in $\Delta a_\text{f}$ the error $\Delta
a_\text{f}^\text{NR}=0.01$ on numerical-relativity results against
which our model is built. We note that $\Delta a_\text{f}\lesssim
0.02$ is compatible with $\sim 1$\% variations of the term
$a_\text{i}M_\text{BH}^2$ appearing in Eq.\,(\ref{eq:model-imp}). If
we interpret this $\sim 1$\% variation as a representation of possible
``glitches'' in the transition from the quasiequilibrium initial data
to the dynamical evolution of the Einstein equations in a numerical
simulation, we see that we are indeed ``inheriting'' a $\sim 0.01$
contribution to $\Delta a_\text{f}$ in building our model against
numerical-relativity results, and that this contribution is at least
comparable to the ones introduced by all other approximations behind
Eqs.\,(\ref{eq:model-imp})-(\ref{eq:fbridge}). Further, we stress once
more that $\Delta a_\text{f}\lesssim 0.02$ is compatible with the
average error, found in~\cite{Barausse:2009uz}, for the BKL model,
which inspired this work. All these conclusions remain valid even when
combining Eqs.\,(\ref{eq:model-imp})-(\ref{eq:fbridge}) with the
method of~\cite{Foucart2012} to calculate $M_\text{b,torus}$.

It is striking that our simple model to determine $a_\text{f}$ paired
with~\cite{Foucart2012} obtains such an excellent agreement with the
fully general-relativistic numerical simulations of BH-NS mergers. One
must always bear in mind, however, that there still are large,
unexplored portions of the parameter space and that this prevents us
from thoroughly testing our approach to determine $a_\text{f}$.

%%%%%%%%%%%%%%%%%%%%%%%%%%%%%%%%%%%%%%%%%%%%%%%%%%%%%%%%%%%%%%%%%%%%%%%%%%%%%%%
\subsection{Testing the final mass predictions}\label{sec:Mftests}
%%%%%%%%%%%%%%%%%%%%%%%%%%%%%%%%%%%%%%%%%%%%%%%%%%%%%%%%%%%%%%%%%%%%%%%%%%%%%%%
So far, we tested only one of the two predictions that our model
enables us to make. In this section we separately test the predictions
for the mass $M_\text{f}$ of the BH remnant, stemming from
Eq.\,(\ref{eq:model-Mf}). According to the ``no-hair'' theorem of
general relativity, the final spin parameter and mass of an
electrically neutral BH are the only two quantities characterizing the
BH itself. A model capable of accurately predicting both $a_\text{f}$
and $M_\text{f}$ would therefore fully describe the BH remnant.

The numerical simulations of BH-NS mergers performed by the
Kyoto-Tokyo group reported
in~\cite{Shibata2012err,Kyutoku2011err,Kyutoku2011} allow us to test
the outcome of Eq.\,(\ref{eq:model-Mf}) and to establish the error
associated with it. In Tables \ref{tab:tests3} and \ref{tab:tests4},
we collect the numerical-relativity data for the mass of the BH
remnant and compare it to our predictions. The first six columns of
the tables follow Tables \ref{tab:tests1} and \ref{tab:tests2},
including the numbering of the simulations appearing in column
one. The seventh and eighth columns provide the relative error on the
remnant masses obtained when comparing the predictions of
Eq.\,(\ref{eq:model-Mf}) to the numerical-relativity
results. Following~\cite{Shibata2012err,Kyutoku2011err,Kyutoku2011},
two forms of the remnant mass are considered: the gravitational mass
$M_\text{f}$, and the irreducible mass
\begin{align}
\label{eq:Mirrf}
M_\text{irr,f} = M_\text{f}\sqrt{\frac{1+\sqrt{1-a_\text{f}^2}}{2}}\,.
\end{align}
Both $M_\text{f}$ and $M_\text{irr,f}$ are divided by the sum $M$ of
the initial gravitational masses $M_\text{BH}$ and $M_\text{NS}$. In
the remaining columns of the table, we give the relative error on the
$l=2$, $m=2$, $n=0$ quasinormal mode (QNM) frequency
$f_{220}^\text{QNM}$, and damping time $\tau_{220}^\text{QNM}$ of the
BH remnant~\cite{Berti:2009kk}. Both $a_\text{f}$ and $M_\text{f}$
must be used to calculate $f_{220}^\text{QNM}$ and
$\tau_{220}^\text{QNM}$, so that $\epsilon(f_{220}^\text{QNM})$ and
$\epsilon(\tau_{220}^\text{QNM})$ give us a sense of how our errors on
the final BH spin parameter and mass propagate. The terms of
comparison for the QNM frequencies and damping times are obtained by
using the final mass and spin parameter values given
in~\cite{Shibata2012err, Kyutoku2011err, Kyutoku2011} and plugging
them in the formulas of~\cite{Berti:2009kk}.

A maximum relative error of $1$\% and $2$\% is found for
$\overline{M}_\text{f}=M_\text{f}/M$ and
$\overline{M}_\text{irr,f}=M_\text{irr,f}/M$, respectively, with the
$2$\% occurring only once. The errors on $f_{220}^\text{QNM}$ and
$\tau_{220}^\text{QNM}$, on the other hand, are at the most $4$\% and
$5$\%, respectively. It is noteworthy that the second contribution in
Eq.\,(\ref{eq:model-Mf}), i.e. the energy loss due to GW emission, is
crucial in obtaining such accurate results: if we do not include it,
the maximum error on $f_{220}^\text{QNM}$, for example, is $9$\%.

%TTTTTTTTTTTTTTTTTTTTTTTTTTTTTTTTTTTTTTTTTTTTTTTTTTTTTTTTTTTTTTTTTTTTT
\begin{table}[!t]
  \caption{\label{tab:tests3} Tests against numerical-relativity results. The first six columns are organized as in Tables \ref{tab:tests1} and \ref{tab:tests2}. The last four columns show the relative error on the BH remnant gravitational mass in units of the system initial mass, $\overline{M}_\text{f}$, on its irreducible mass in units of the system initial mass, $\overline{M}_\text{irr,f}$, and on its $l=2$, $m=2$, $n=0$ quasinormal mode frequency, $f_{220}^\text{QNM}$, and damping time, $\tau_{220}^\text{QNM}$.}
  \begin{tabular}{l@{\hspace{0.15cm}}c@{\hspace{0.15cm}}c@{\hspace{0.15cm}}c@{\hspace{0.15cm}}c@{\hspace{0.15cm}}c@{\hspace{0.1cm}}c@{\hspace{0.1cm}}c@{\hspace{0.1cm}}c@{\hspace{0.1cm}}c}
    \toprule[1.pt]
    \toprule[1.pt]
    \addlinespace[0.3em]
    & Ref. & EOS & $C$ & $Q$ & $a_\text{i}$ & $\epsilon(\overline{M}_\text{f})$ & $\epsilon(\overline{M}_\text{irr,f})$ & $\epsilon(f_{220}^\text{QNM})$ & $\epsilon(\tau_{220}^\text{QNM})$\\
    \addlinespace[0.2em]
    \midrule[1.pt]
    \addlinespace[0.2em]
    5  & \cite{Shibata2012err} & $\Gamma=2$ & $0.145$ & $2$ & $0$ & $0.00$ & $0.01$ & $0.03$ & $0.02$\\
    6  & \cite{Shibata2012err} & $\Gamma=2$ & $0.145$ & $3$ & $0$ & $0.00$ & $0.00$ & $0.01$ & $0.00$\\
    7  & \cite{Shibata2012err} & $\Gamma=2$ & $0.145$ & $4$ & $0$ & $0.00$ & $0.00$ & $0.00$ & $0.00$\\
    8  & \cite{Shibata2012err} & $\Gamma=2$ & $0.145$ & $5$ & $0$ & $0.00$ & $0.00$ & $0.01$ & $0.00$\\
    9  & \cite{Shibata2012err} & $\Gamma=2$ & $0.160$ & $2$ & $0$ & $0.00$ & $0.01$ & $0.02$ & $0.01$\\
    10 & \cite{Shibata2012err} & $\Gamma=2$ & $0.160$ & $3$ & $0$ & $0.00$ & $0.00$ & $0.01$ & $0.00$\\
    11 & \cite{Shibata2012err} & $\Gamma=2$ & $0.178$ & $2$ & $0$ & $0.00$ & $0.01$ & $0.01$ & $0.00$\\
    12 & \cite{Shibata2012err} & $\Gamma=2$ & $0.178$ & $3$ & $0$ & $0.01$ & $0.00$ & $0.00$ & $0.01$\\
    \midrule
    20 & \cite{Kyutoku2011err} & 2H-135   & $0.131$ & $2$ & $0$ & $0.00$ & $0.00$ & $0.01$ & $0.00$\\
    21 & \cite{Kyutoku2011err} & H-135    & $0.162$ & $2$ & $0$ & $0.00$ & $0.00$ & $0.01$ & $0.00$\\
    22 & \cite{Kyutoku2011err} & HB-135   & $0.172$ & $2$ & $0$ & $0.00$ & $0.01$ & $0.01$ & $0.00$\\
    23 & \cite{Kyutoku2011err} & HBs-135  & $0.172$ & $2$ & $0$ & $0.00$ & $0.01$ & $0.01$ & $0.00$\\
    24 & \cite{Kyutoku2011err} & HBss-135 & $0.174$ & $2$ & $0$ & $0.00$ & $0.01$ & $0.01$ & $0.00$\\
    25 & \cite{Kyutoku2011err} & B-135    & $0.182$ & $2$ & $0$ & $0.01$ & $0.00$ & $0.01$ & $0.01$\\
    26 & \cite{Kyutoku2011err} & Bs-135   & $0.185$ & $2$ & $0$ & $0.01$ & $0.00$ & $0.00$ & $0.01$\\
    27 & \cite{Kyutoku2011err} & Bss-135  & $0.194$ & $2$ & $0$ & $0.01$ & $0.00$ & $0.00$ & $0.02$\\
    28 & \cite{Kyutoku2011err} & 2H-135   & $0.131$ & $3$ & $0$ & $0.00$ & $0.00$ & $0.00$ & $0.00$\\
    29 & \cite{Kyutoku2011err} & H-135    & $0.162$ & $3$ & $0$ & $0.01$ & $0.00$ & $0.01$ & $0.01$\\
    30 & \cite{Kyutoku2011err} & HB-135   & $0.172$ & $3$ & $0$ & $0.01$ & $0.00$ & $0.01$ & $0.01$\\
    31 & \cite{Kyutoku2011err} & B-135    & $0.182$ & $3$ & $0$ & $0.01$ & $0.01$ & $0.00$ & $0.01$\\
    32 & \cite{Kyutoku2011err} & 2H-12    & $0.118$ & $2$ & $0$ & $0.00$ & $0.00$ & $0.00$ & $0.00$\\
    33 & \cite{Kyutoku2011err} & H-12     & $0.145$ & $2$ & $0$ & $0.00$ & $0.00$ & $0.01$ & $0.01$\\
    34 & \cite{Kyutoku2011err} & HB-12    & $0.153$ & $2$ & $0$ & $0.00$ & $0.00$ & $0.01$ & $0.00$\\
    35 & \cite{Kyutoku2011err} & B-12     & $0.161$ & $2$ & $0$ & $0.00$ & $0.00$ & $0.01$ & $0.00$\\
    36 & \cite{Kyutoku2011err} & HB-12    & $0.153$ & $3$ & $0$ & $0.00$ & $0.00$ & $0.00$ & $0.00$\\
    37 & \cite{Kyutoku2011err} & B-12     & $0.161$ & $3$ & $0$ & $0.01$ & $0.01$ & $0.01$ & $0.01$\\
    \bottomrule[1.pt]
    \bottomrule[1.pt]
  \end{tabular}
\end{table}
%TTTTTTTTTTTTTTTTTTTTTTTTTTTTTTTTTTTTTTTTTTTTTTTTTTTTTTTTTTTTTTTTTTTTT

%TTTTTTTTTTTTTTTTTTTTTTTTTTTTTTTTTTTTTTTTTTTTTTTTTTTTTTTTTTTTTTTTTTTTT
\begin{table}[!t]
  \caption{\label{tab:tests4} Same as Table \ref{tab:tests3}.}
  \resizebox{\columnwidth}{!}{%
  \begin{tabular}{l@{\hspace{0.12cm}}c@{\hspace{0.12cm}}c@{\hspace{0.14cm}}c@{\hspace{0.14cm}}c@{\hspace{0.13cm}}d{2.2}@{\hspace{-0.2cm}}d{2.2}@{\hspace{-0.2cm}}d{2.2}@{\hspace{-0.2cm}}d{2.2}@{\hspace{-0.2cm}}d{2.2}}
    \toprule[1.pt]
    \toprule[1.pt]
    \addlinespace[0.3em]
    & Ref. & EOS & $C$ & $Q$ & \multicolumn{1}{c}{$a_\text{i}$} & \multicolumn{1}{c}{$\epsilon(\overline{M}_\text{f})$} & \multicolumn{1}{c}{$\epsilon(\overline{M}_\text{irr,f})$} & \multicolumn{1}{c}{$\epsilon(f_{220}^\text{QNM})$} & \multicolumn{1}{c}{$\epsilon(\tau_{220}^\text{QNM})$}\\
    \addlinespace[0.2em]
    \midrule[1.pt]
    \addlinespace[0.2em]
    38 & \cite{Kyutoku2011} & 2H-135   & $0.131$ & $2$ &  0.75 & 0.01 & 0.02 & 0.04 & 0.04\\
    39 & \cite{Kyutoku2011} & 1.5H-135 & $0.146$ & $2$ &  0.75 & 0.00 & 0.01 & 0.00 & 0.00\\
    40 & \cite{Kyutoku2011} & H-135    & $0.162$ & $2$ &  0.75 & 0.00 & 0.00 & 0.02 & 0.03\\
    41 & \cite{Kyutoku2011} & HB-135   & $0.172$ & $2$ &  0.75 & 0.00 & 0.01 & 0.02 & 0.03\\
    42 & \cite{Kyutoku2011} & B-135    & $0.182$ & $2$ &  0.75 & 0.00 & 0.01 & 0.02 & 0.03\\
    43 & \cite{Kyutoku2011} & 2H-135   & $0.131$ & $2$ &  0.5  & 0.00 & 0.00 & 0.01 & 0.01\\
    44 & \cite{Kyutoku2011} & 1.5H-135 & $0.146$ & $2$ &  0.5  & 0.00 & 0.00 & 0.01 & 0.01\\
    45 & \cite{Kyutoku2011} & H-135    & $0.162$ & $2$ &  0.5  & 0.00 & 0.00 & 0.00 & 0.00\\
    46 & \cite{Kyutoku2011} & HB-135   & $0.172$ & $2$ &  0.5  & 0.00 & 0.00 & 0.01 & 0.01\\
    47 & \cite{Kyutoku2011} & B-135    & $0.182$ & $2$ &  0.5  & 0.00 & 0.01 & 0.01 & 0.01\\
    48 & \cite{Kyutoku2011} & 2H-135   & $0.131$ & $2$ & -0.5  & 0.00 & 0.00 & 0.01 & 0.01\\
    49 & \cite{Kyutoku2011} & H-135    & $0.162$ & $2$ & -0.5  & 0.00 & 0.00 & 0.01 & 0.00\\
    50 & \cite{Kyutoku2011} & HB-135   & $0.172$ & $2$ & -0.5  & 0.00 & 0.00 & 0.00 & 0.00\\
    51 & \cite{Kyutoku2011} & B-135    & $0.182$ & $2$ & -0.5  & 0.00 & 0.00 & 0.01 & 0.01\\
    52 & \cite{Kyutoku2011} & 2H-135   & $0.131$ & $3$ &  0.75 & 0.00 & 0.01 & 0.02 & 0.01\\
    53 & \cite{Kyutoku2011} & 1.5H-135 & $0.146$ & $3$ &  0.75 & 0.00 & 0.01 & 0.01 & 0.02\\
    54 & \cite{Kyutoku2011} & H-135    & $0.162$ & $3$ &  0.75 & 0.00 & 0.01 & 0.02 & 0.04\\
    55 & \cite{Kyutoku2011} & HB-135   & $0.172$ & $3$ &  0.75 & 0.00 & 0.00 & 0.01 & 0.02\\
    56 & \cite{Kyutoku2011} & B-135    & $0.182$ & $3$ &  0.75 & 0.01 & 0.00 & 0.02 & 0.05\\
    57 & \cite{Kyutoku2011} & 2H-135   & $0.131$ & $3$ &  0.5  & 0.00 & 0.00 & 0.01 & 0.01\\
    58 & \cite{Kyutoku2011} & 1.5H-135 & $0.146$ & $3$ &  0.5  & 0.00 & 0.00 & 0.01 & 0.01\\
    59 & \cite{Kyutoku2011} & H-135    & $0.162$ & $3$ &  0.5  & 0.00 & 0.00 & 0.01 & 0.01\\
    60 & \cite{Kyutoku2011} & HB-135   & $0.172$ & $3$ &  0.5  & 0.01 & 0.00 & 0.01 & 0.01\\
    61 & \cite{Kyutoku2011} & B-135    & $0.182$ & $3$ &  0.5  & 0.01 & 0.00 & 0.00 & 0.02\\
    62 & \cite{Kyutoku2011} & HB-135   & $0.172$ & $3$ & -0.5  & 0.00 & 0.00 & 0.01 & 0.00\\
    63 & \cite{Kyutoku2011} & 2H-135   & $0.131$ & $4$ &  0.75 & 0.00 & 0.01 & 0.01 & 0.01\\
    64 & \cite{Kyutoku2011} & H-135    & $0.162$ & $4$ &  0.75 & 0.01 & 0.00 & 0.02 & 0.04\\
    65 & \cite{Kyutoku2011} & HB-135   & $0.172$ & $4$ &  0.75 & 0.01 & 0.00 & 0.01 & 0.02\\
    66 & \cite{Kyutoku2011} & B-135    & $0.182$ & $4$ &  0.75 & 0.01 & 0.01 & 0.00 & 0.02\\
    67 & \cite{Kyutoku2011} & 2H-135   & $0.131$ & $4$ &  0.5  & 0.00 & 0.00 & 0.01 & 0.01\\
    68 & \cite{Kyutoku2011} & H-135    & $0.162$ & $4$ &  0.5  & 0.01 & 0.01 & 0.01 & 0.01\\
    69 & \cite{Kyutoku2011} & HB-135   & $0.172$ & $4$ &  0.5  & 0.01 & 0.01 & 0.01 & 0.01\\
    70 & \cite{Kyutoku2011} & B-135    & $0.182$ & $4$ &  0.5  & 0.01 & 0.01 & 0.01 & 0.01\\
    71 & \cite{Kyutoku2011} & 2H-135   & $0.131$ & $5$ &  0.75 & 0.00 & 0.01 & 0.02 & 0.03\\
    72 & \cite{Kyutoku2011} & H-135    & $0.162$ & $5$ &  0.75 & 0.01 & 0.00 & 0.01 & 0.02\\
    73 & \cite{Kyutoku2011} & HB-135   & $0.172$ & $5$ &  0.75 & 0.01 & 0.00 & 0.00 & 0.02\\
    74 & \cite{Kyutoku2011} & B-135    & $0.182$ & $5$ &  0.75 & 0.01 & 0.00 & 0.00 & 0.03\\
    \bottomrule[1.pt]
    \bottomrule[1.pt]
  \end{tabular}}
\end{table}
%TTTTTTTTTTTTTTTTTTTTTTTTTTTTTTTTTTTTTTTTTTTTTTTTTTTTTTTTTTTTTTTTTTTTT

\begin{figure*}
  \includegraphics[width=8cm]{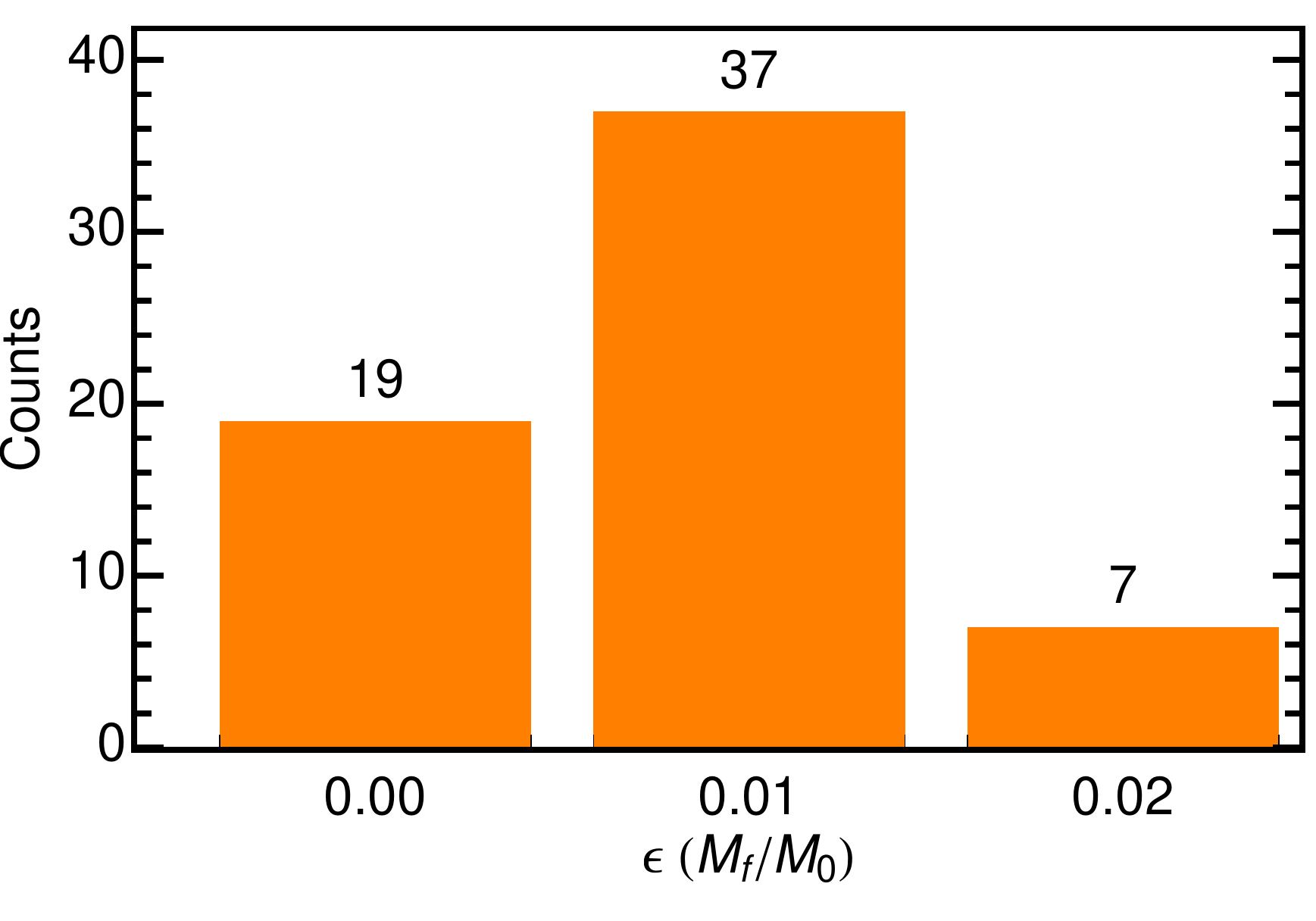}
  \hskip 0.5cm
  \includegraphics[width=8cm]{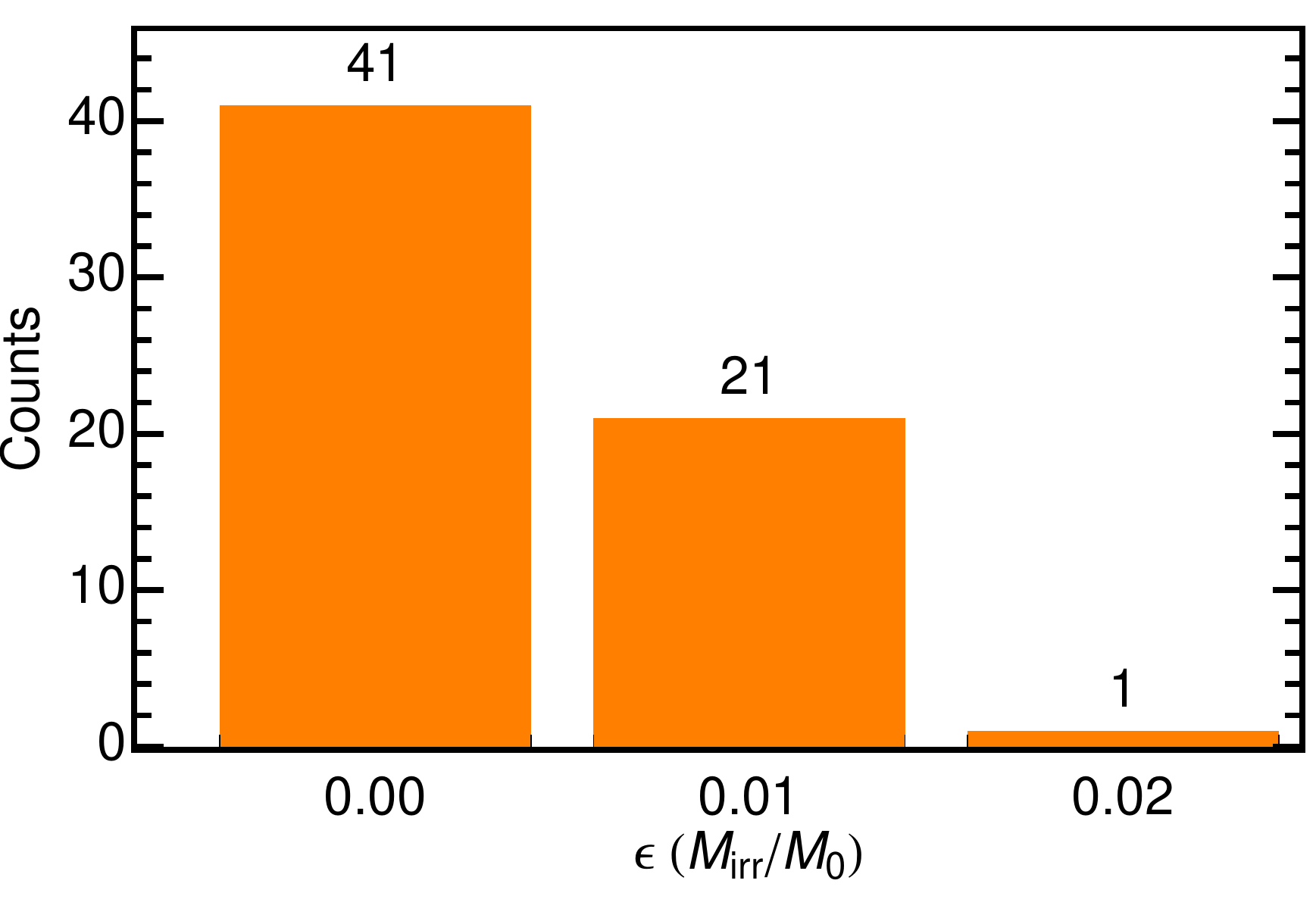}
  \vskip 0.75cm
  \includegraphics[width=8cm]{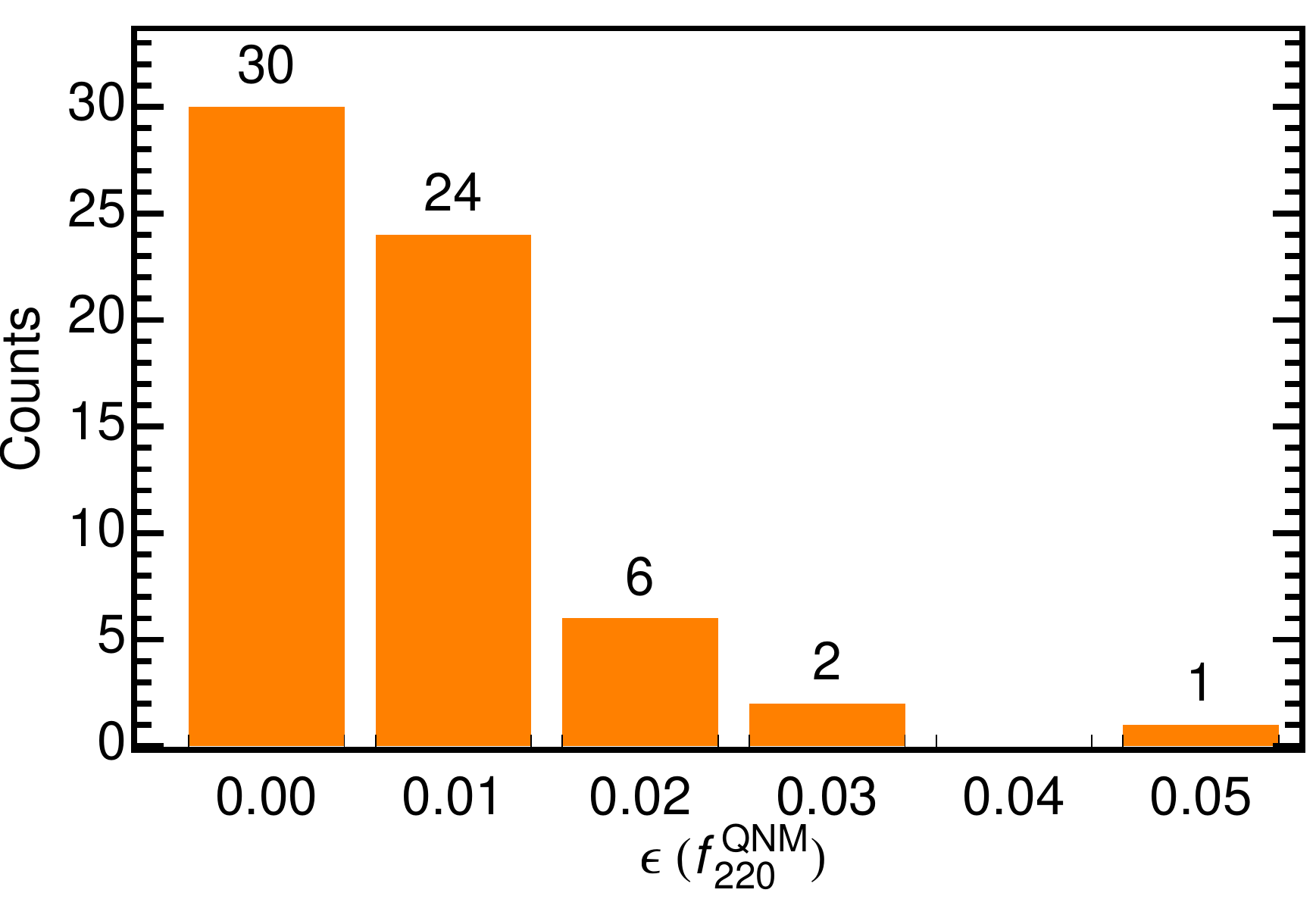}
  \hskip 0.5cm
  \includegraphics[width=8cm]{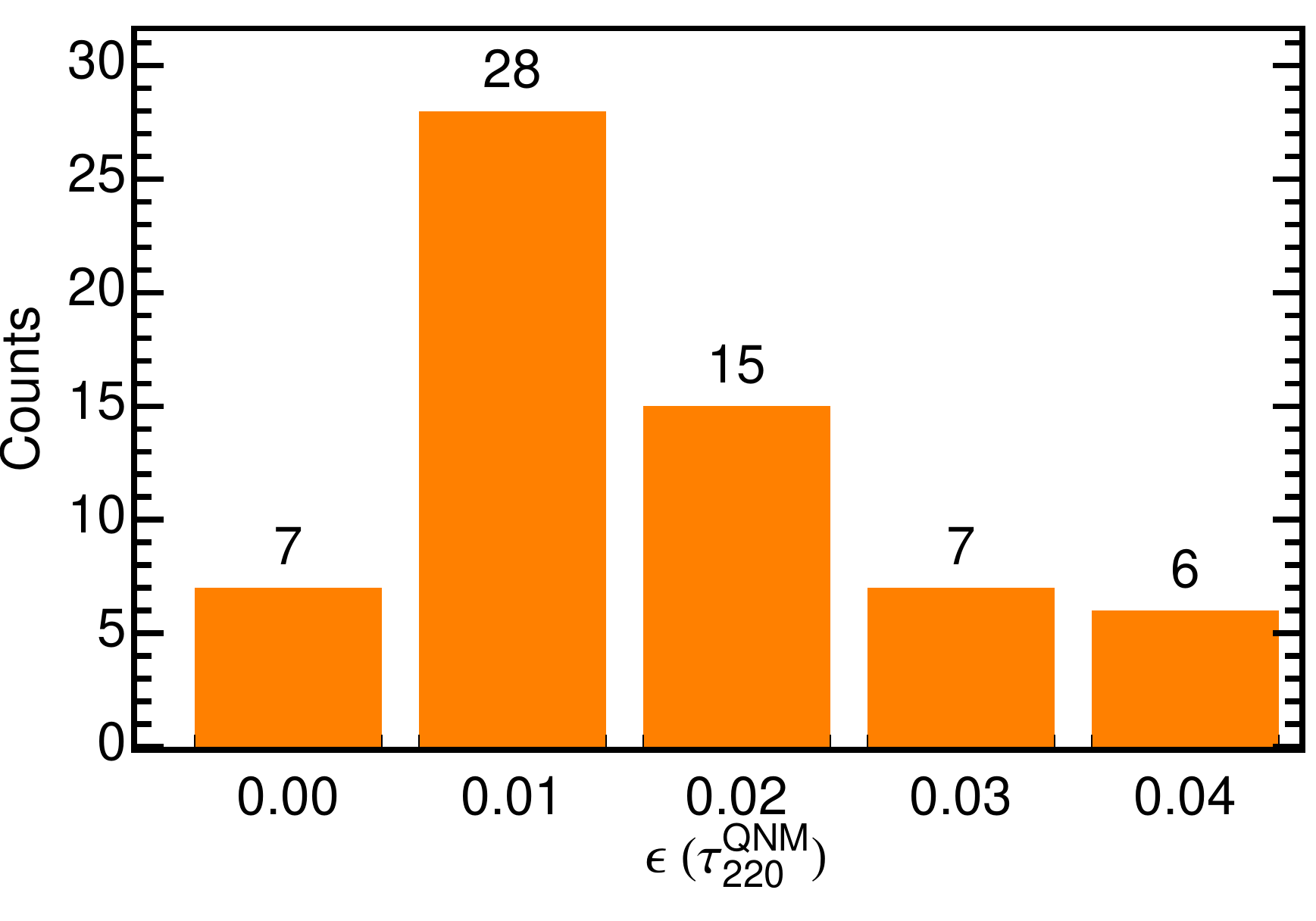}
  \caption{\label{FIG:HistoMQNMErrors} Distribution of the relative
    errors $\epsilon(M_\text{f}/M)$, $\epsilon(M_\text{irr,f}/M)$,
    $\epsilon(f^\text{QNM}_{220})$, and
    $\epsilon(\tau^\text{QNM}_{220})$ obtained from
    Eqs.\,(\ref{eq:model-Mf}), (\ref{eq:model-imp})-(\ref{eq:fbridge})
    in combination with~\cite{Foucart2012} for test cases in Tables
    \ref{tab:tests3} and \ref{tab:tests4} for which information on the
    BH remnant mass is provided in the references.}
\end{figure*}

If we use input from the model of~\cite{Foucart2012} and repeat these
tests on $\overline{M}_\text{f}$, $\overline{M}_\text{irr,f}$,
$f_{220}^\text{QNM}$, and $\tau_{220}^\text{QNM}$, the maximum errors
we obtain are $2$\%, $3$\%, $5$\%, and $4$\%, respectively. The panels
of Fig.\,\ref{FIG:HistoMQNMErrors} show the distributions of the
relative errors $\epsilon(\overline{M}_\text{f})$,
$\epsilon(\overline{M}_\text{irr,f})$, $\epsilon(f_{220}^\text{QNM})$,
and $\epsilon(\tau_{220}^\text{QNM})$ obtained when using
Eqs.\,(\ref{eq:model-Mf}), (\ref{eq:model-imp})-(\ref{eq:fbridge}) in
combination with~\cite{Foucart2012}. As for the tests performed with
the numerical-relativity values of $M_\text{b,torus}$, these
distributions are peaked around $\sim 0.00-0.01$ and errors higher
than $2$\% are rare. We stress once more that large portions of the
parameter space of BH-NS binaries are currently unexplored, thus
preventing us from testing our approach thoroughly.

%%%%%%%%%%%%%%%%%%%%%%%%%%%%%%%%%%%%%%%%%%%%%%%%%%%%%%%%%%%%%%%%%%%%%%%%%%%%%%%
\section{Results}\label{sec:results}
%%%%%%%%%%%%%%%%%%%%%%%%%%%%%%%%%%%%%%%%%%%%%%%%%%%%%%%%%%%%%%%%%%%%%%%%%%%%%%%
We now review the main results obtained by systematically exploring
the space of parameters of BH-NS systems using the model described so
far. More specifically, we vary
\ben[~~~~(i)]
\item the initial spin parameter of the BH, $a_\text{i}$, reaching a
  maximum value of $0.99$;
\item the binary mass ratio, $Q$, between $2$ and $10$;
\item the NS mass, between $1.2M_\odot$ and $2.0M_\odot$, compatibly
  with the measurement reported in~\cite{Antoniadis2013};
\item the NS compactness. In particular, we use the WFF1
  EOS~\cite{Wiringa88} and the PS EOS~\cite{PandharipandeSmith:1975}
  as representatives of the softest and stiffest possible EOS,
  yielding the most and least compact NSs, respectively. Thus, for a
  given NS mass we consider the compactness of a NS governed by the
  WFF1 EOS and the one of a NS described by the PS EOS. We also quote
  results for the APR2 EOS~\cite{Akmal1997,Akmal1998a} since this is
  the most complete nuclear many-body study to date.
\een
%

%FFFFFFFFFFFFFFFFFFFFFFFFFFFFFFFFFFFFFFFFFFFFFFFFFFFFFFFFFFFFFFFFFFFFF
%FFFFFFFFFFFFFFFFFFFFFFFFFFFFFFFFFFFFFFFFFFFFFFFFFFFFFFFFFFFFFFFFFFFFF

\noindent The choices regarding the EOS are discussed in detail in the
Appendix.

We analyze the behavior of $a_\text{f}$ in the relevant space of
parameters, examining its maximum possible value; we further compare
the outcome of BH-NS mergers and BH-BH mergers in terms of the
$l=m=2$, $n=0$ quasinormal mode frequency and show that the comparison
is EOS dependent. When comparing to binary black holes, we apply the
fitting formula of~\cite{Barausse:2009uz} to determine the final spin
parameter of the their remnants, and we neglect the last term in
Eq.\,(\ref{eq:model-Mf}) to estimate the mass of their
remnants.\footnote{This is how the BH remnant properties are
  determined in~\cite{Santamaria2010}.} More accurate predictions for
$M_\text{f}$ are possible for BH-BH binaries,
e.g.~\cite{Barausse2012b}; if we were to rely on them, however, we
would be comparing predictions with a different degree of accuracy,
thus mixing the physical consequences of replacing the lower mass BH
of a binary BH with a NS to effects due to the different precision
underlying the predictions under comparison.

%%%%%%%%%%%%%%%%%%%%%%%%%%%%%%%%%%%%%%%%%%%%%%%%%%%%%%%%%%%%%%%%%%%%%%%%%%%%%%%
\subsection{Maximum final spin parameter}\label{sec:maxaf}
%%%%%%%%%%%%%%%%%%%%%%%%%%%%%%%%%%%%%%%%%%%%%%%%%%%%%%%%%%%%%%%%%%%%%%%%%%%%%%%

An important aspect to investigate when studying the final spin of the
BH remnant of compact binary mergers is its maximum value. According
to the cosmic censorship conjecture, the spin parameter of a BH cannot
exceed unity~\cite{Penrose79}. Indirect support to the conjecture was
provided by the recent numerical-relativity simulations of BH-NS
mergers~\cite{Kyutoku2011}. The extrapolation of the results of the
numerical simulations to the case of an extremely spinning BH with
$a_\text{i}=1$ (merging with an irrotational NS) yielded
$a_\text{f}\sim 0.98$. It was suggested that simulations with mass
ratio higher than $Q=4$ and (nearly) extremal initial BH spin should
be performed in order to assess whether $a_\text{f}\lesssim 0.98(<1)$
is a universal bound for BH-NS binary mergers or not.

%FFFFFFFFFFFFFFFFFFFFFFFFFFFFFFFFFFFFFFFFFFFFFFFFFFFFFFFFFFFFFFFFFFFFF
\begin{figure*}
  \includegraphics[width=8cm]{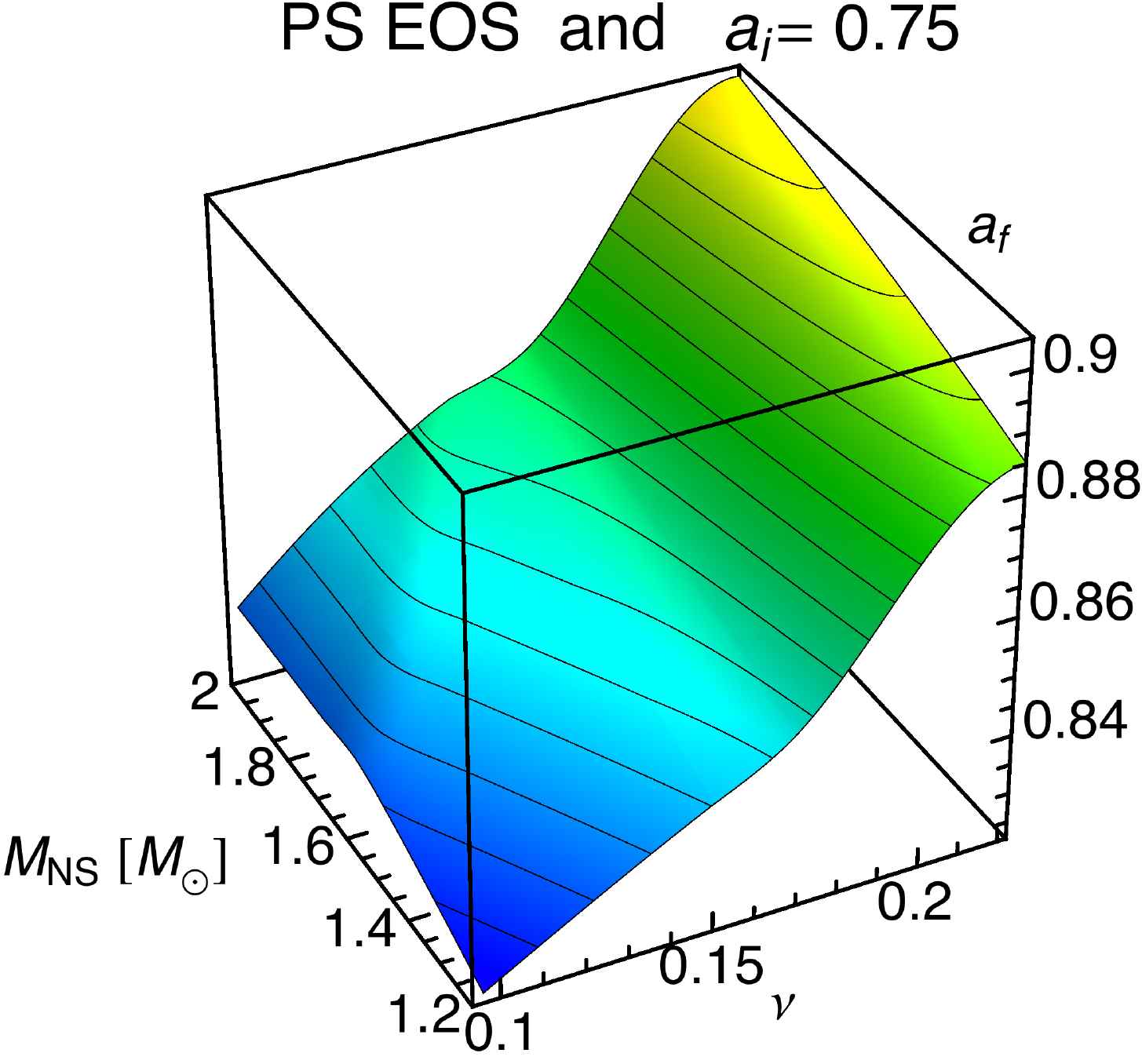}
  \includegraphics[width=8cm]{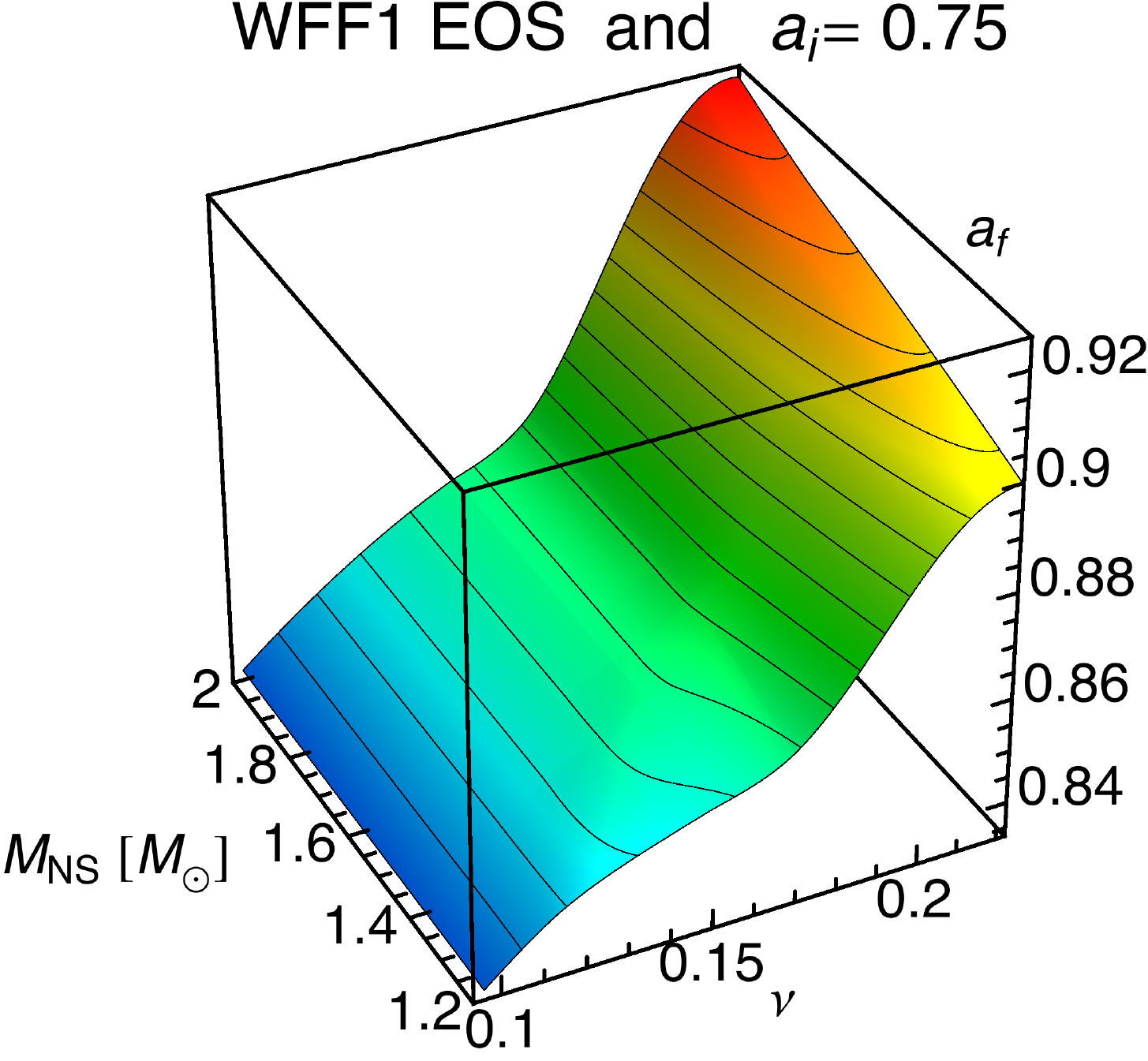}
  \caption{\label{FIG:a075} (Color online). The final spin parameter
    of BH-NS systems is shown as a function of the NS mass and the
    symmetric mass ratio; the initial spin parameter of the BH is set
    to $0.75$. Results in the left/right panel refer to the PS/WFF1
    EOS.}
\end{figure*}
%FFFFFFFFFFFFFFFFFFFFFFFFFFFFFFFFFFFFFFFFFFFFFFFFFFFFFFFFFFFFFFFFFFFFF

%FFFFFFFFFFFFFFFFFFFFFFFFFFFFFFFFFFFFFFFFFFFFFFFFFFFFFFFFFFFFFFFFFFFFF
\begin{figure*}[]
  \includegraphics[width=5.3cm]{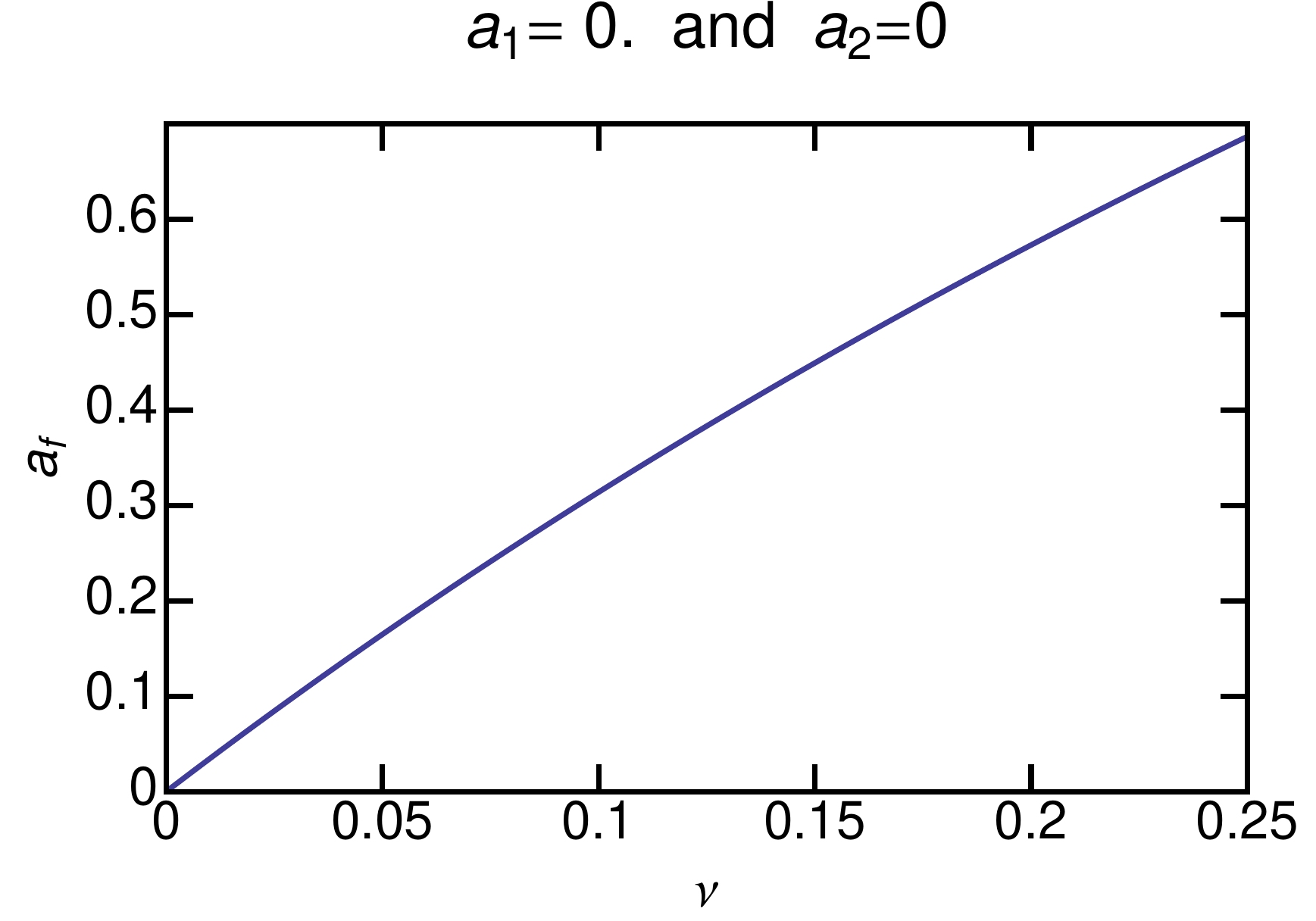}
  \hskip 0.5cm
  \includegraphics[width=5.3cm]{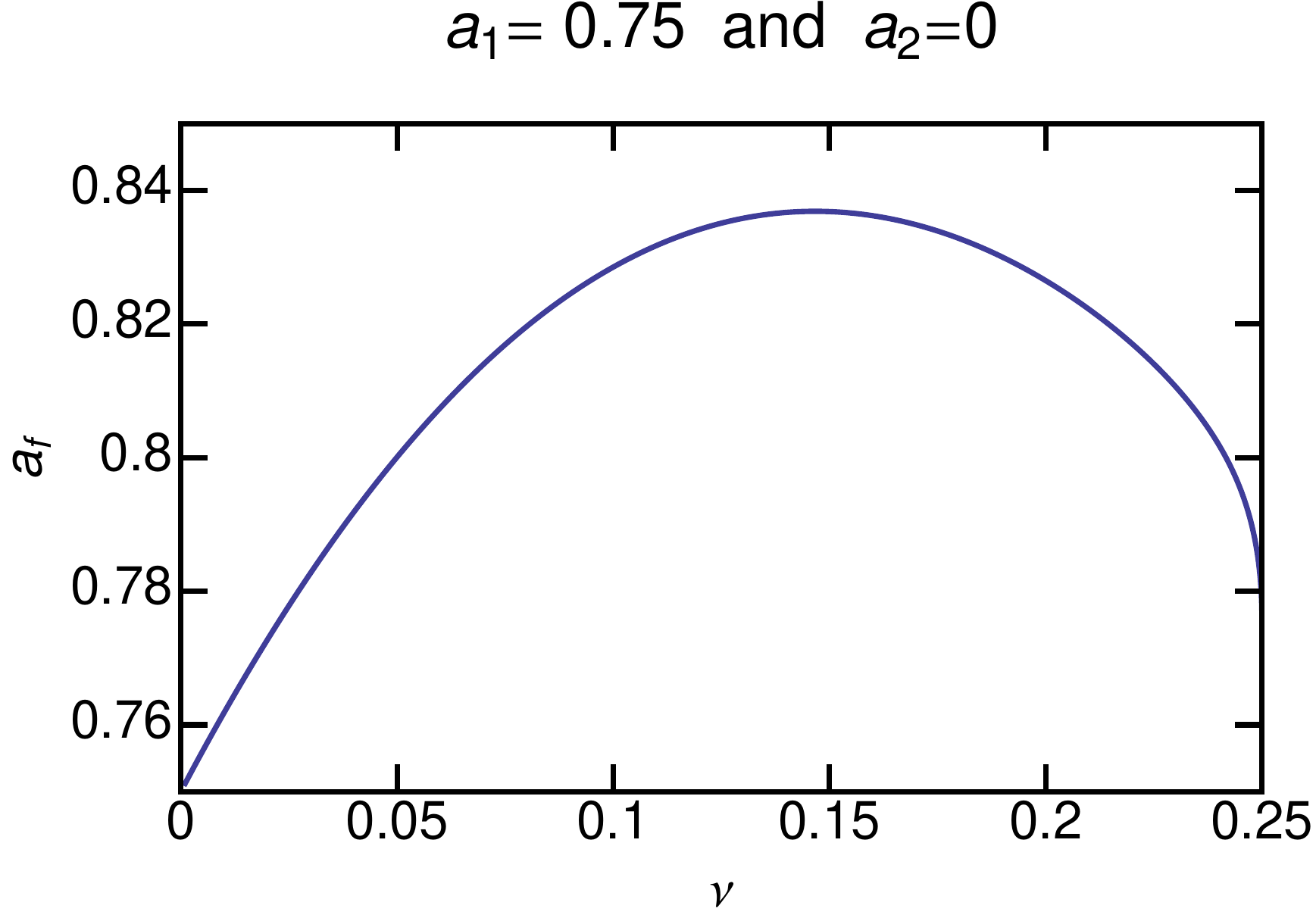}
  \hskip 0.5cm
  \includegraphics[width=5.3cm]{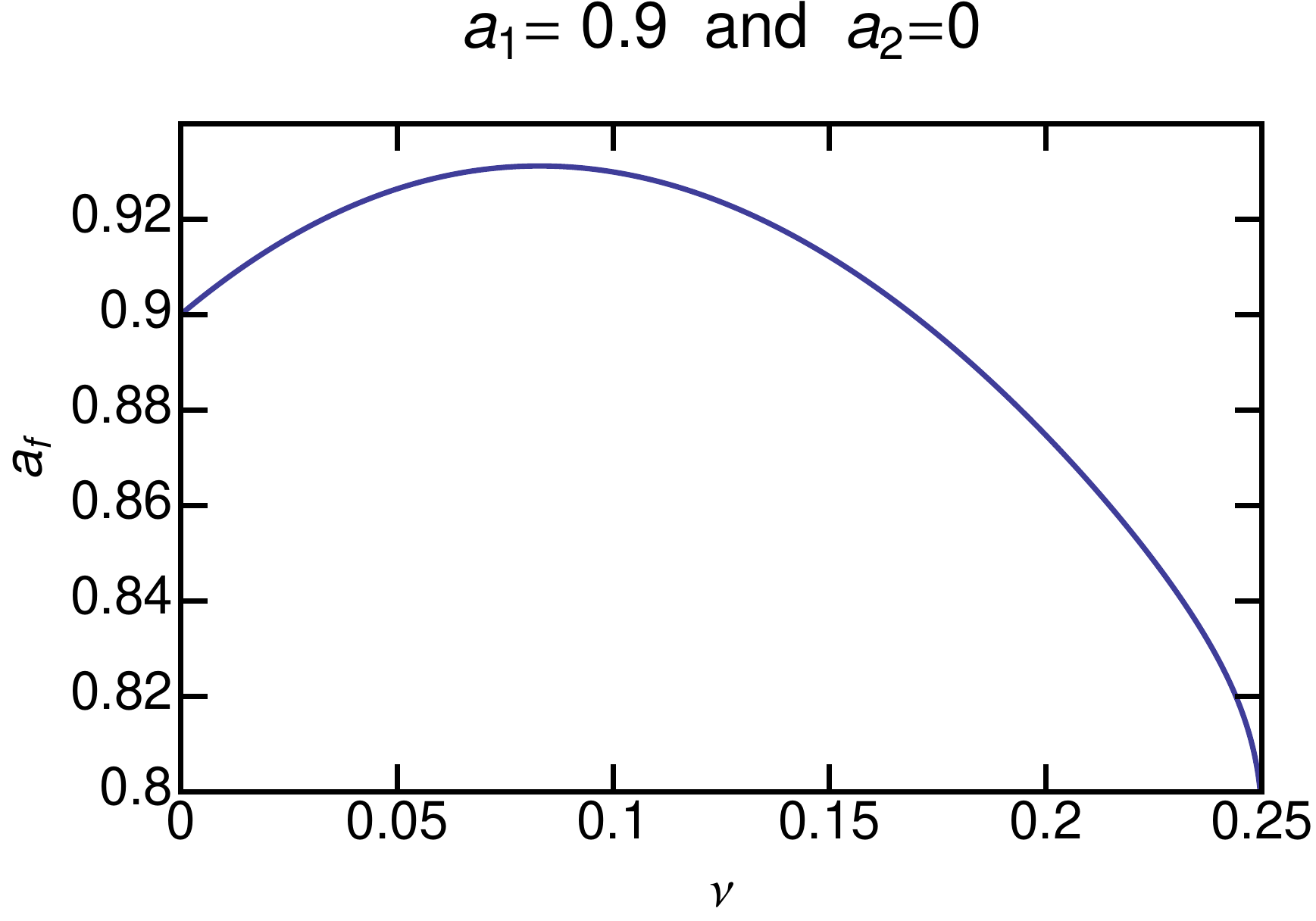}
  \caption{\label{FIG:BBHs} (Color online). Final spin parameter for
    binaries with a primary spinning BH and a secondary nonspinning
    BH. From left to right, the initial spin parameter of the spinning
    BH is set to $0$, $0.75$, and $0.9$. The curves were obtained with
    the the fitting formula of~\cite{Barausse:2009uz}.}
\end{figure*}
%FFFFFFFFFFFFFFFFFFFFFFFFFFFFFFFFFFFFFFFFFFFFFFFFFFFFFFFFFFFFFFFFFFFFF

Our model does not predict the formation of overspinning
($a_\text{f}>1$) BHs for BH-NS binaries with an extremal initial BH
spin and any symmetric mass ratio. We notice that, all else being
fixed, the softer the EOS, the higher the final spin parameter
$a_\text{f}$. We thus suggest performing fully general-relativistic
numerical simulations of systems with (nearly) extremal initial BH
spin parameter and a soft NS EOS to assess the bound on $a_\text{f}<1$
for BH-NS binary mergers.

To determine the maximum final spin parameter, we consider our data
and extrapolate it to $a_\text{i}=1$. We perform the extrapolations on
two different sets of data: in one case we use all our data, i.e. with
$a_\text{i}$ up to $0.99$, whereas in the other we consider only
$a_\text{i}\leq 0.9$. This allows us to cross-check our predictions
obtained within the untested region of the parameter space
$0.9<a_\text{i}\leq 1$, thus making our conclusions more robust. The
highest final spin parameters we obtain are for $Q=10$, or $\nu\simeq
0.083$. The WFF1, PS, and APR2 EOS all yield a maximum final spin
parameter $a_\text{f}=1.00$, compatible with the $a_\text{f}\sim 0.98$
bound pinpointed in~\cite{Kyutoku2011}. Even though we previously
determined that our predictions have an error $\Delta
a_\text{f}\lesssim 0.02$ (see discussion in Sec.\,\ref{sec:model}), we
believe it is worth mentioning that we find $\max a_\text{f}=0.997$,
an ``empirical'' result that is very close to Thorne's limit of
$0.998$~\cite{Thorne1974}.

%%%%%%%%%%%%%%%%%%%%%%%%%%%%%%%%%%%%%%%%%%%%%%%%%%%%%%%%%%%%%%%%%%%%%%%%%%%%%%%
\subsection{Dependence of the BH remnant on the NS
  EOS}\label{sec:QNMs}
%%%%%%%%%%%%%%%%%%%%%%%%%%%%%%%%%%%%%%%%%%%%%%%%%%%%%%%%%%%%%%%%%%%%%%%%%%%%%%%
%FFFFFFFFFFFFFFFFFFFFFFFFFFFFFFFFFFFFFFFFFFFFFFFFFFFFFFFFFFFFFFFFFFFFF
\begin{figure*}[]
  \includegraphics[width=17cm]{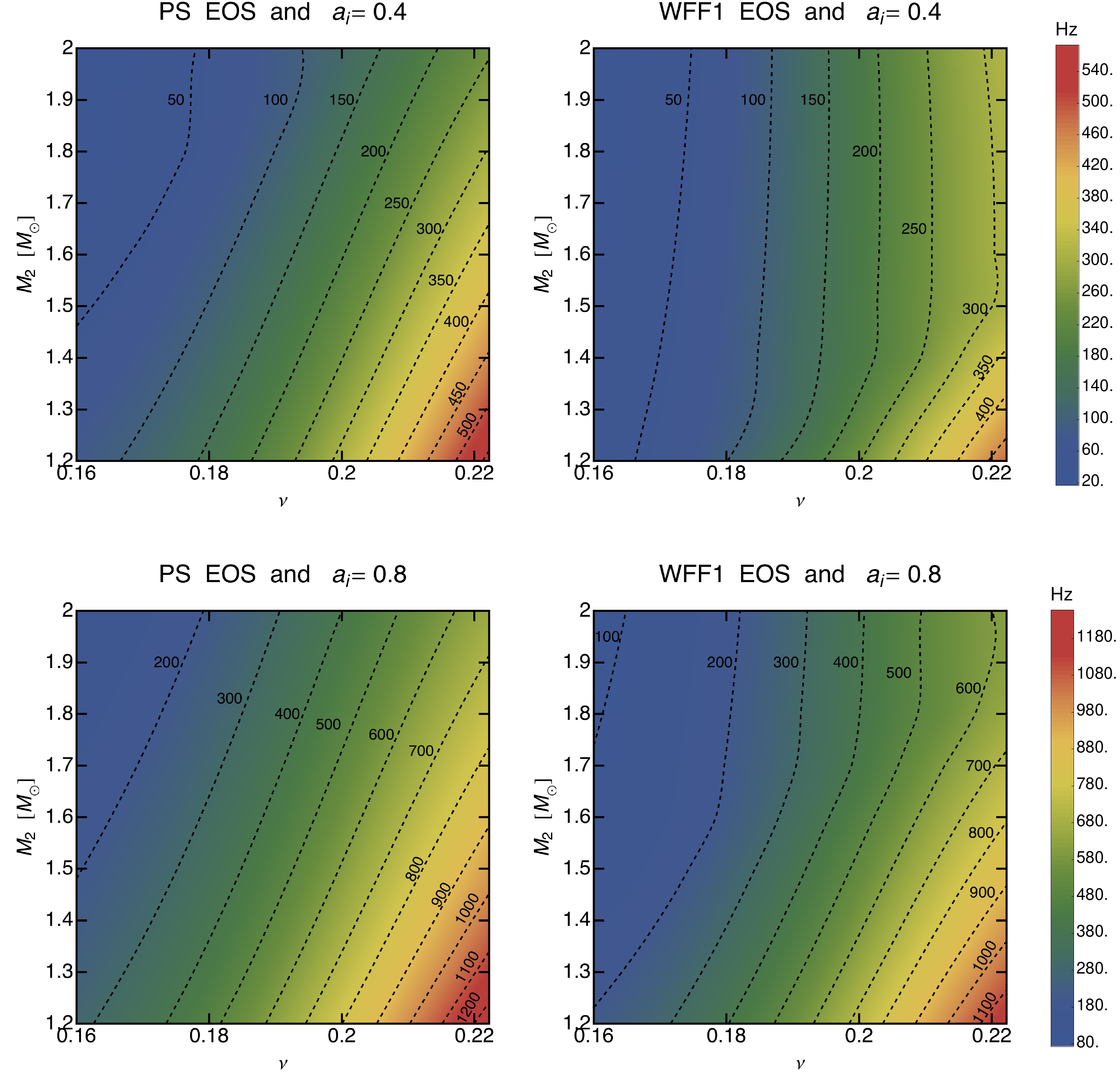}
  \caption{\label{FIG:BHBH_QNM} (Color online). Difference, in Hz,
    between the $l=m=2$, $n=0$ quasinormal mode frequency,
    $f_{220}^\text{QNM}$, of the BH remnant of a BH-NS merger and of a
    BH-BH merger with the same secondary mass $M_2$, symmetric mass
    ratio $\nu$, and initial spin parameters. The NS and secondary BH
    initial spin is always set to zero. The primary BH initial spin
    parameter is $0.4$ and $0.8$ in the top and bottom panels,
    respectively.}
\end{figure*}
%FFFFFFFFFFFFFFFFFFFFFFFFFFFFFFFFFFFFFFFFFFFFFFFFFFFFFFFFFFFFFFFFFFFFF

The main feature that appears when comparing results for different
EOSs is that final spin parameter of the BH remnant, $a_\text{f}$, can
depend on the EOS of the NS in the mixed binary progenitor (all else
being fixed). This happens because different EOSs yield different
torus masses. An example of this EOS dependence is provided in
Fig.\,\ref{FIG:a075}, where binaries with $a_\text{i}=0.75$ and two
possible EOSs, the WFF1 and the PS, are considered. The EOS dependence
of $a_\text{f}$ may be better understood by carefully examining the
case of BH-BH binaries. Figure \ref{FIG:BBHs} shows that, given a
binary with a nonspinning secondary BH and a primary BH with initial
spin parameter $a_\text{i}$, there is a specific symmetric mass ratio
that yields the maximum $a_\text{f}$: its value varies monotonically
from $0.25$ to $0$ as the $a_\text{i}$ runs from $0$ to $1$. More
specifically, the first panel shows that for nonspinning binary BHs
higher values of $a_\text{f}$ are favored by high symmetric mass
ratios (i.e. it is ``easier'' to spin up a Schwarzschild BH with a
mass comparable to the one of the BH itself), while this is not true
in the other two panels, in which the primary BH is rotating. In the
case of BH-NS systems, as one varies $\nu$ and $M_\text{NS}$, the BH
mass changes along with the mass accreting onto the BH: in the case of
disruptive mergers, the latter depends on the EOS and this explains
why the spin of the BH remnant depends on the EOS of the NS in the
progenitor for BH-NS disruptive mergers.

Having shown that $a_\text{f}$ may depend on the NS EOS and bearing in
mind that $M_\text{f}$ may too,\footnote{This is more straightforward
  to comprehend: NSs differing only for the EOS, differ in compactness
  and may thus accrete different amounts of matter in disruptive
  mergers with a BH, e.g.~\cite{Kyutoku2011}.} this means that
information about the NS EOS is ``coded'' in the properties of the BH
remnant. This in turn implies that the QNM spectrum of the BH remnant
may (1) be affected by the EOS and (2) deviate from the BH-BH
behavior. In Fig.\,\ref{FIG:BHBH_QNM} we compare the BH remnant of
BH-NS mergers to the BH remnant of BH-BH mergers. We show the
difference between the QNM frequency $f_{220}^\text{QNM}$ of the BH
remnant of a BH-NS merger and of a BH-BH merger with the same
secondary mass, symmetric mass ratio, and initial spin parameters. The
final spin parameter and mass of the binary BHs are determined using
the method of~\cite{Barausse:2009uz} and Eq.\,(\ref{eq:model-Mf})
without the last term, respectively. All QNM frequencies are
calculated using the fitting formula of~\cite{Berti:2009kk}.

Our results confirm that the mixed binaries we expect to see the most
of, i.e.~those with $\nu\sim 0.11$~\cite{Belczynski07}, do indeed
behave like binary BHs in terms of GW emission during the ringdown
epoch: this is positive for template design and GW detection. In
Fig.\,\ref{FIG:BHBH_QNM} we show our results in the region above
$\nu=0.16$ for binaries with $a_\text{i}=0.4$ and $a_\text{i}=0.8$,
and we contrast the PS EOS with the WFF1 EOS. We find that for
$\nu\sim 0.16$ (or $Q\sim 4$) the BH remnant QNM frequency
$f_{220}^\text{QNM}$ deviates from its BH-BH binary value by $\lesssim
100$Hz for both soft and stiff EOSs, unless the initial BH spin
parameter is particularly high and the NS EOS is very stiff (bottom,
left panel of Fig.\,\ref{FIG:BHBH_QNM}). The NS thus leaves a (small)
``trace'' in the QNM frequency: for binaries with a BH with moderate
to high spin and a symmetric mass ratio $\nu\lesssim 0.16$, one could
in principle determine whether the source of a detected GW coalescence
signal was a BH-BH or a BH-NS binary by separately analysing the
inspiral and the ringdown epochs. The former epoch would be identical
for a mixed binary and a BH binary with the same physical parameters,
because tidal deformations of the NS in a mixed binary with $\nu
\lesssim 0.16$ are not expected to significantly alter the inspiral
epoch of the GW signal~\cite{Pannarale2011}. Looking at the ringdown
epoch would therefore complement the idea of pinning down the presence
of the NS from the inspiral. Constraining the NS EOS in the region of
the parameter space around $\nu\sim 0.16$ by measuring
$f_{220}^\text{QNM}$ appears, instead, to be difficult.

For systems with high symmetric mass ratios, on the other hand, the
difference in $f_{220}^\text{QNM}$ may be high, suggesting the
interesting prospect of constraining the NS EOS through the
measurement of the properties of the BH remnant. If we now focus on
the high $\nu$ region of the panels in Fig.\,\ref{FIG:BHBH_QNM}, we see
that the deviations in $f_{220}^\text{QNM}$ from the BH-BH case are
particularly evident for $\nu \gtrsim 0.2$ and a high initial BH spin
parameter. This and all other features of the bottom panels in
Fig.\,\ref{FIG:BHBH_QNM} may be compared to those of the panels
Fig.\,\ref{FIG:a075}, showing that they are ``inherited'' from the
behavior of $a_\text{f}$. At high $\nu$'s and $a_\text{i}$'s, the
difference between the $f_{220}^\text{QNM}$ of the BH remnant of a
BH-NS merger and the one of a BH-BH merger ranges from $\sim 600$Hz to
$\sim 1200$Hz. The NS EOS therefore leaves an imprint on the QNM
frequency of the BH remnant, although one should bear in mind that we
are comparing two (extreme) EOSs that are on opposite ends in terms of
stiffness, and this makes the differences between the left and the
right panels of Fig.\,\ref{FIG:BHBH_QNM} particularly prominent. A
comparison between results for the WFF1 EOS and the APR2 EOS, which
yield NSs relatively similar in terms of compactness (see
Fig.\,\ref{FIG:MvsC}), tells us that in order to be able to properly
discriminate between similar candidate nuclear EOSs, one would need to
be able to perform measurements of $f_{220}^\text{QNM}$ with a
precision the order of $\sim 10$Hz.

Being able to perform measurements of $f_{220}^\text{QNM}$ for the BH
remnant of BH-NS mergers requires that the QNM itself is excited
during the coalescence. If this happens, $f_{220}^\text{QNM}$
influences the value of the cutoff frequency $f_\text{cut}$ of the GW
spectrum of the mixed binary
coalescence.\footnote{See~\cite{Kyutoku2011} for explanations on what
  determines $f_\text{cut}$ and for examples of gravitational waveform
  spectra.} The extrapolation of the results of the
numerical-relativity simulations reported in~\cite{Kyutoku2011} shows,
in particular, that $f_\text{cut}\simeq f_{220}^\text{QNM}$ for
$C\gtrsim 0.18$ in mixed binaries with $\nu\simeq 0.139$ (or $Q=5$)
and $a_\text{i}=0.75$, or for $C\gtrsim 0.19$ in binaries when
$a_\text{i}=0$ and $\nu\simeq 0.22$ ($Q=2$) or $\nu\simeq 0.1875$
($Q=3$) and $a_\text{i}=0.5$. An increase in $a_\text{i}$ corresponds
to an increase in the lower bound on $C$ that allows for
$f_\text{cut}\simeq f_{220}^\text{QNM}$ to happen at a given mass
ratio; on the other hand, the higher the mass of the BH at a given
$a_\text{i}$ and $C$, the closer $f_\text{cut}$ will be to
$f_{220}^\text{QNM}$. The $\sim 0.19$ threshold on the NS compactness
encountered above corresponds to $M_\text{NS}\gtrsim 1.33M_\odot$ for
the WFF1 EOS, to $M_\text{NS}\gtrsim 1.48M_\odot$ for the APR2 EOS,
and to $M_\text{NS}\gtrsim 1.93M_\odot$ for the PS EOS. We thus see
that there is the virtual possibility of constraining the NS EOS with
the measurement of the gravitational radiation emitted by those
binaries for which $f_\text{cut}\simeq f_{220}^\text{QNM}$. This
scenario, as said, concerns NSs with high compactness and would thus
provide constraints for soft EOSs. We note that the observation of
tidal effects in the phase of the gravitational radiation emitted
during the inspiral~\cite{Hinderer09, Pannarale2011} and in the cutoff
frequency when $f_\text{cut}<f_{220}^\text{QNM}$~\cite{Vallisneri00,
  Kyutoku2011} favors placing constraints on stiff EOSs, so that
measurements in $f_\text{cut}\sim f_{220}^\text{QNM}$ scenarios would
be complementary.

%%%%%%%%%%%%%%%%%%%%%%%%%%%%%%%%%%%%%%%%%%%%%%%%%%%%%%%%%%%%%%%%%%%%%%%%%%%%%%%
\section{Conclusions and Remarks}\label{sec:conclusions}
%%%%%%%%%%%%%%%%%%%%%%%%%%%%%%%%%%%%%%%%%%%%%%%%%%%%%%%%%%%%%%%%%%%%%%%%%%%%%%%
In this paper we presented a model for predicting the final spin
parameter, $a_\text{f}$, and mass, $M_\text{f}$, of the BH remnant of
BH-NS coalescing binaries in quasicircular orbits, with initial BH
spin of arbitrary magnitude and parallel to the orbital angular
momentum, with arbitrary mass ratio, and with arbitrary NS mass and
cold, barotropic equation of state. The parameter space just outlined
could in principle be investigated entirely with numerical-relativity
simulations; in practice, however, the process would be very time and
resource consuming, because simulations are still very expensive in
terms of computational costs.

Our starting point was the phenomenological model of Buonanno, Kidder,
and Lehner for the final spin of binary BH
mergers~\cite{Buonanno:07b}, which we modified to account for (1)
energy loss via gravitational wave emission during the inspiral and
(2) the possible formation of an accretion torus in the case of
disruptive mergers. We tested our model by comparing its predictions
to the recent numerical-relativity simulation results available in the
literature. We were able to achieve good agreement down to a mass
ratio of $M_\text{BH}/M_\text{NS}=2$, albeit introducing an additional
ingredient in the formulation of the model for
$2<M_\text{BH}/M_\text{NS}<4$ which is currently poorly
constrained. We obtained an absolute error on $a_\text{f}$ of $0.02$,
which is compatible with the one of the BKL
approach~\cite{Buonanno:07b,Barausse:2009uz}. For the final
gravitational and irreducible (normalized) masses of the BH remnant,
$M_\text{f}$ and $M_\text{irr,f}$, we found a relative error of
$1$\%. These errors then propagate in the calculation of the $l=m=2$,
$n=0$ quasinormal mode frequency $f_{220}^\text{QNM}$ and damping time
$\tau_{220}^\text{QNM}$ of the remnant BH, and they yield maximum
relative errors of $4$\% and $5$\%, respectively. These relative
errors are, however, safely $\leq 2$\% in the vast majority of test
cases. Combining this method with input for the torus remnant mass
from the two-parameter model, fitted to existing numerical results,
recently reported by Foucart in~\cite{Foucart2012}, the error $\Delta
a_\text{f}\simeq 0.02$ is preserved and so is the behavior of the
relative errors on $M_\text{f}$, $M_\text{irr,f}$,
$f_{220}^\text{QNM}$, and $\tau_{220}^\text{QNM}$ (see
Figs.\,\ref{FIG:AbsErrorsToy} and \ref{FIG:HistoMQNMErrors}).

The tests we performed against the available numerical-relativity
results were successful, especially considering the limitations of our
simple approach. This implies that the outcome of the complicated
merger dynamics of BH-NS binaries may be understood in fairly simple
terms, at least when the BH initial spin and orbital angular momentum
directions are parallel and the inspiral orbit is
quasicircular. Equations (\ref{eq:model-Mf}),
(\ref{eq:model-imp})-(\ref{eq:fbridge}) presented here along with the
fit of Foucart to the torus remnant mass~\cite{Foucart2012} constitute
an easy-to-use analytical model that describes the remnant of BH-NS
mergers. Notwithstanding the good performance of the model in the
tests, its predictions should be taken with ``a grain of salt,'' as
large portions of the parameter space of BH-NS binaries are currently
unexplored, hampering a thorough test of our approach.

The approach presented and tested in the first part of this work was
then employed to span the space of parameters consisting of the binary
(symmetric) mass ratio, the BH initial spin, the NS mass, and the NS
equation of state. This enabled us to gain a sense of the behavior of
the properties of the BH remnant in this fourfold space of parameters
and to pinpoint some interesting aspects which, we believe, deserve
being verified and studied in conclusive, quantitative terms with the
tools of numerical relativity. The following is a summary of our main
results:
\ben[~~~~(i)]
\item We obtained a maximum final spin parameter equal to $1.00$ for
  the WFF1, APR2, and PS nuclear equations of state, when using the
  mass ratio $M_\text{BH}/M_\text{NS}=10$. $M_\text{BH}/M_\text{NS}=2$
  yields instead a maximum final spin parameter of $0.99$. Given their
  absolute error of $0.02$, these predictions are compatible with the
  $0.98$ maximum found in~\cite{Kyutoku2011} and provide indirect
  support to the cosmic censorship conjecture~\cite{Penrose79}.
\item We discussed the dependence of $a_\text{f}$ and $M_\text{f}$ on
  the NS EOS, claiming that the EOS may leave an imprint on the BH
  remnant. The quasinormal mode frequency $f_{220}^\text{QNM}$ of the
  BH remnant, which depends on $a_\text{f}$ and $M_\text{f}$ alone,
  could thus be used to constrain the NS EOS
  (Fig.\,\ref{FIG:BHBH_QNM}). Deviations from the BH-BH values of
  $f_{220}^\text{QNM}$ for symmetric mass ratios $\gtrsim 0.2$, with
  maximum deviations between $\sim 600$Hz and $\sim 1200$Hz. The
  excitation of the QNM oscillations does not occur for all mixed
  binary mergers, but it is likely to appear in the spectrum of the
  emitted gravitational radiation in the form of a cutoff frequency
  $f_\text{cut}\simeq f_{220}^\text{QNM}$ for systems with fairly
  compact NSs, i.e.~for soft EOSs~\cite{Kyutoku2011}. The possibility
  of constraining the EOS by measuring $f_\text{cut}\simeq
  f_{220}^\text{QNM}$ seems therefore complementary to other ideas for
  posing EOS constraints by means of GW detection, in that these favor
  constraints on stiff EOSs~\cite{Vallisneri00, Hinderer09,
    Pannarale2011, Kyutoku2011}. High-frequency gravitational waves
  from coalescing binaries may thus turn out to be, once more, very
  promising in terms of the NS EOS~\cite{Bauswein2012}.
\een

Future applications of the approach presented in this paper may be to
exploit the predicted values of $a_\text{f}$ and $M_\text{f}$ to (1)
provide the QNM frequencies to be used in the construction of hybrid
waveforms for BH-NS systems~\cite{Kyutoku2011,Lackey2012}, (2) develop
phenomenological waveforms for BH-NS systems, (3) study time-frequency
characteristics of the emitted radiation~\cite{Hanna2008}, and (4)
build backgrounds for perturbative approaches to the study of the
postmerger epoch. Two main extensions are, instead, foreseeable and
consist in allowing for a more general initial state for the
binary. One may investigate dropping the assumption that (1) the
neutron star is initially irrotational and that (2) the initial spin
angular momentum of the black hole and the orbital angular momentum
are parallel. In the former case, one would have to add a spin angular
momentum contribution from the neutron star to the numerator of
Eq.\,(\ref{eq:model-imp}) and allow for a fraction of this angular
momentum to possibly be dissipated prior to the disruption/plunge of
the star. A major obstacle, however, would be that, since there are no
numerical-relativity simulations with a nonirrotational neutron star
initial state, we lack a model to predict $M_\text{b,torus}$ when the
neutron star is initially spinning. As this quantity appears in
Eq.\,(\ref{eq:model-imp}), being able to predict it would be a
fundamental gap to fill in, prior to extending the model discussed in
this paper. Regarding cases with nonparallel black hole spin angular
momentum and orbital angular momentum, the ISCO-related expressions
appearing in Eq.\,(\ref{eq:model-imp}) would have to be
generalized. Input in this direction has recently started to emerge
from the numerical-relativity community via the first simulations of
tilted BH-NS mergers, e.g.~\cite{Foucart2010}, that would serve as
test cases. In general, and independently of extending the model, it
is important to keep testing the model as more numerical simulations
are published in the literature, possibly constraining the ansatz of
Eq.\,(\ref{eq:fbridge}) and improving the model of~\cite{Foucart2012},
on which the entire approach relies.

In concluding this work, we would like to stimulate the BH-NS
numerical-relativity community to continue investigating different
parameter configurations, as this would allow us to better constrain
the current version of our model [see the discussion following
Eq.\,(\ref{eq:model-imp})]. Investigations on more generic initial
spin configurations would also be helpful~\cite{Foucart2010}, as they
would allow us to look into extending our approach.

%%%%%%%%%%%%%%%%%%%%%%%%%%%%%%%%%%%%%%%%%%%%%%%%%%%%%%%%%%%%%%%%%%%%%%%%%%%%%%%
\section*{Acknowledgments}
%%%%%%%%%%%%%%%%%%%%%%%%%%%%%%%%%%%%%%%%%%%%%%%%%%%%%%%%%%%%%%%%%%%%%%%%%%%%%%%
It is a pleasure to thank Emanuele Berti, John Friedman, Philipp
M\"osta, Frank Ohme, Luciano Rezzolla, and James Ryan for useful
discussions and comments, and Tim Dietrich for carefully reading the
manuscript. The author is grateful to the authors
of~\cite{Foucart2013a} for testing the model discussed in this paper
against their new results, and validating it further. This work was
supported in part by the DFG Grant SFB/Transregio~7.

\appendix
%%%%%%%%%%%%%%%%%%%%%%%%%%%%%%%%%%%%%%%%%%%%%%%%%%%%%%%%%%%%%%%%%%%%%%%%%%%%%%%
\section*{APPENDIX: Equations of state}\label{app:EOS}
%%%%%%%%%%%%%%%%%%%%%%%%%%%%%%%%%%%%%%%%%%%%%%%%%%%%%%%%%%%%%%%%%%%%%%%%%%%%%%%
NSs are the most compact objects known, lacking an event horizon. The
densities in the interior of these stars are expected to exceed the
equilibrium density of nuclear matter ($\rho_s\simeq 2.7\cdot
10^{14}\,$g/cm$^{-3}$), so their macroscopic properties, e.g. mass and
radius, and the internal composition of their cores depend on the
nature of strong interactions in dense matter and reflect (different
aspects of) the dense matter EOS. Our knowledge about the behavior of
matter at such exceptionally high densities, however, is still
currently limited. As far as the composition is concerned, for
example, several dense matter models predict that --- in addition to
nucleons, electrons, and muons --- exotica in the form of hyperons, a
Bose condensate of mesons, or deconfined quark matter eventually
appear at supranuclear densities~\cite{Lattimer07}.

\begin{figure}[!t]
  \begin{center}
    \includegraphics[width=8.0cm,angle=-0]{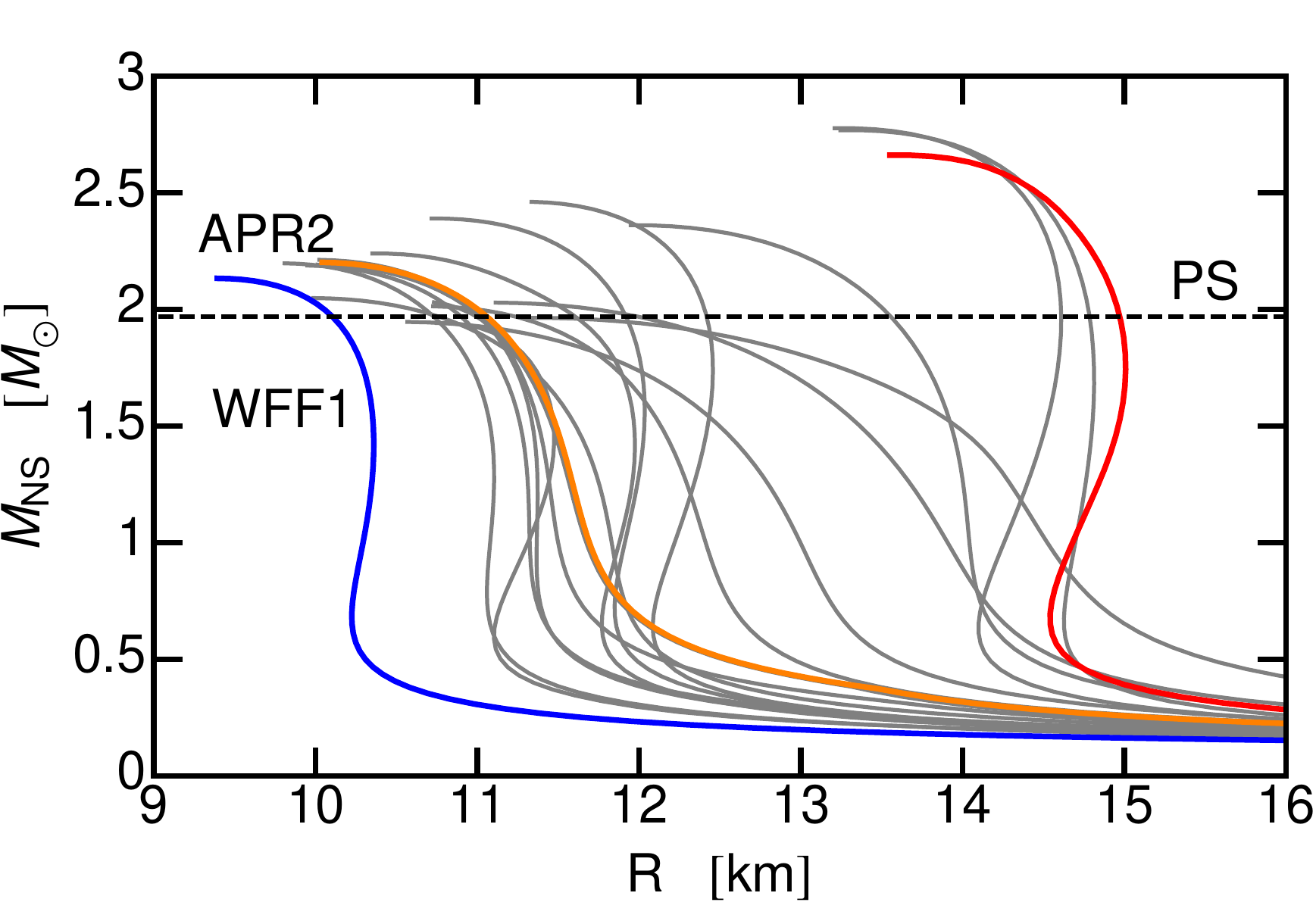}
    \caption{\label{FIG:MvsR} (Color online). NS equilibrium
      sequences in the radius-mass plane for several equations of
      state. The sequences shown are compatible with the recent
      measurement $M_\text{NS}=(2.01\pm 0.04)M_\odot$ (horizontal,
      dashed line). Results for the WFF1/PS EOS, which yields the
      most/least compact NSs, are shown in blue/red. The APR2 sequence
      is shown in orange, while sequences obtained with other
      equations of state are shown with thinner, continuous gray
      lines.}
  \end{center}
\end{figure}

\begin{figure}[!t]
  \begin{center}
    \includegraphics[width=8.0cm,angle=-0]{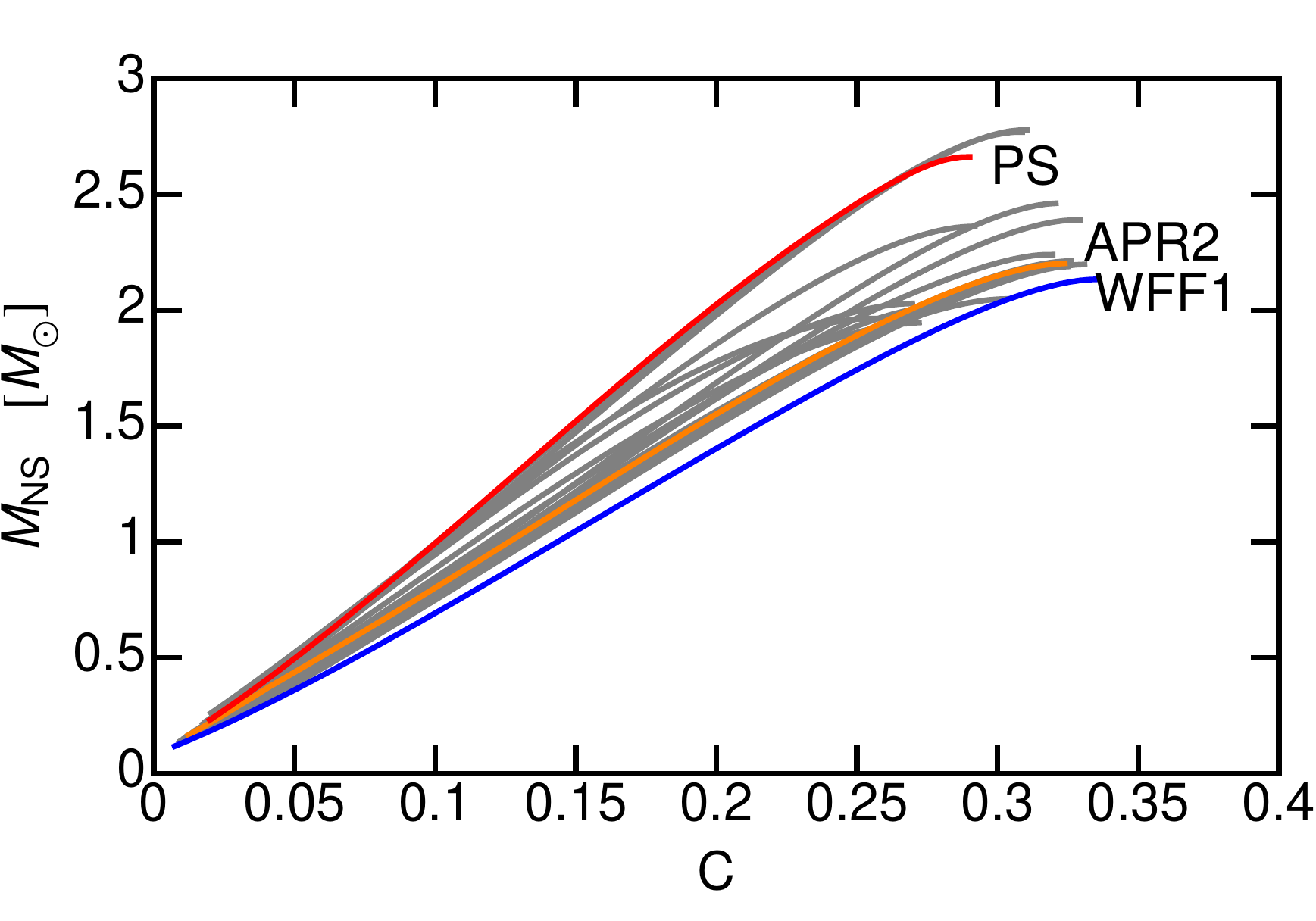}
    \caption{\label{FIG:MvsC} (Color online). Same as
      Figure~\ref{FIG:MvsR}, but in the compactness-mass plane.}
  \end{center}
\end{figure}

An intense investigation to determine the EOS of dense matter was
performed throughout the
years~\cite{Alford2001,Lattimer2006,Lattimer07}. The recent
measurement of a NS with mass $M_\text{NS}=(2.01\pm 0.04)M_\odot$
ruled out several equations of state proposed over
time~\cite{Antoniadis2013}. NS equilibrium sequences for
EOSs\footnote{We do not consider strange quark matter equations of
  state.}  compatible with such measurement are shown in the
radius-mass plane in Fig.\,\ref{FIG:MvsR}. In order to assess the
impact of the EOS on the BH remnant of BH-NS mergers, we pick the two
EOSs that yield the smallest and the largest NS radii for any given NS
mass between \hbox{$\sim 1M_\odot$} and $\sim 2.1M_\odot$. We dub
these two EOSs WFF1 and PS, respectively, because of their NS core
description~\cite{Wiringa88,PandharipandeSmith:1975}. The NS
equilibrium sequences they yield are shown in blue and red in
Fig.\,\ref{FIG:MvsR}. As is obvious from the graph, the WFF1 and the
PS EOS bracket all other equations of state in the relevant NS mass
range\footnote{The theoretical minimum mass for a proto-neutron star
  is $1.1M_\odot-1.2M_\odot$~\cite{Goussard1998}.} $1.0\lesssim
M_\text{NS}/M_\odot\lesssim 2.1$. The sequences of
Fig.\,\ref{FIG:MvsR} are displayed in the compactness-mass plane in
Fig.\,\ref{FIG:MvsC}.

%FFFFFFFFFFFFFFFFFFFFFFFFFFFFFFFFFFFFFFFFFFFFFFFFFFFFFFFFFFFFFFFFFFFFF
%FFFFFFFFFFFFFFFFFFFFFFFFFFFFFFFFFFFFFFFFFFFFFFFFFFFFFFFFFFFFFFFFFFFFF

%FFFFFFFFFFFFFFFFFFFFFFFFFFFFFFFFFFFFFFFFFFFFFFFFFFFFFFFFFFFFFFFFFFFFF
%FFFFFFFFFFFFFFFFFFFFFFFFFFFFFFFFFFFFFFFFFFFFFFFFFFFFFFFFFFFFFFFFFFFFF

For both the WFF1 and the PS EOS, we use the same description of
matter in the outer layers of the NS:
\ben[~~~~(i)]
\item for densities in the interval starting at the neutron drip
  density $4\cdot 10^{11}\,$g/cm$^3$ and ending at $2\cdot
  10^{14}\,$g/cm$^3$, the Pethick-Ravenhall-Lorenz (PRL)
  EOS~\cite{Pethick1995} is used;
\item for the crust layer in the density interval $(10^7-4\cdot
  10^{11})\,$g/cm$^3$, the Baym-Pethick-Sutherland (BPS)
  EOS~\cite{Baym71b} is adopted;
\item and, finally, for densities lower than $10^7\,$g/cm$^3$ the BPS
  EOS is extrapolated.
\een
The two EOSs differ at densities above $2\cdot 10^{14}\,$g/cm$^3$: for
the NS core we use what are strictly speaking the WFF1 EOS
of~\cite{Wiringa88} and the ``liquid'' version of the PS EOS
of~\cite{PandharipandeSmith:1975}. The WFF1 EOS for dense nuclear
matter is based on a many-body Hamiltonian built with the Argonne
$v_{14}$ two-nucleon potential and the Urbana VII three-nucleon
potential; calculations are performed with a variational method. The
PS EOS, instead, considers neutron-only matter with $\pi^0$
condensates; the $\pi^0$ relativistic field is not treated explicitly
but is instead replaced by an equivalent two-body potential;
calculations are performed using a constrained variational
method. Both WFF1 and PS are dated and have been superseded by more
modern models and calculation techniques; however, they serve our
purpose of considering extremely compact and extremely large NSs,
respectively, to explore the space of parameters of BH-NS binaries.

In addition to the WFF1 and PS cases, we also discuss results obtained
for the APR2 EOS~\cite{Akmal1997,Akmal1998a}, which is used as it
represents the most complete nuclear many-body study to date and
special-relativistic corrections were progressively incorporated in
it. APR2 is based on the Argonne $v_{18}$ two-nucleon potential, the
Urbana IX three-nucleon potential, and the $\delta v_b$ boost; it is
supported by current astrophysical~\cite{Steiner2010} and nuclear
physics constraints~\cite{Hebeler2010}.

%%%%%%%%%%%%%%%%%%%%%%%%%%%%%%%%%%%%%%%%%%%%%%%%%%%%%%%%%%%%%%%%%%%%%%%%%%%%%% 
%%%%%%%%%%%%%%%%%%%%%%%%%%%%%%%%%%%%%%%%%%%%%%%%%%%%%%%%%%%%%%%%%%%%%%%%%%%%%% 
\bibliographystyle{apsrev4-1-noeprint}
\bibliography{aeireferences}
%%%%%%%%%%%%%%%%%%%%%%%%%%%%%%%%%%%%%%%%%%%%%%%%%%%%%%%%%%%%%%%%%%%%%%%%%%%%%%%
%%%%%%%%%%%%%%%%%%%%%%%%%%%%%%%%%%%%%%%%%%%%%%%%%%%%%%%%%%%%%%%%%%%%%%%%%%%%%%%
\end{document}